\documentclass[aps,prd,reprint,longbibliography,nofootinbib,superscriptaddress]{revtex4-2}
\usepackage{lmodern}
\usepackage[T1]{fontenc}
\usepackage{amsmath}
\usepackage{amssymb}

\usepackage{url}
\usepackage{multirow}
\usepackage{caption}
\usepackage[labelformat=simple]{subcaption}

\usepackage[nodayofweek]{datetime}
\usepackage{enumerate}
\usepackage{xspace}
\usepackage{xcolor}
\usepackage{graphicx}
\usepackage[pdftex,hidelinks]{hyperref}
\usepackage{bookmark}
\usepackage[compat=1.1.0]{tikz-feynman}
\usepackage{placeins}

\usepackage[capitalise]{cleveref}

\usepackage{import}
\usepackage{tabularx, booktabs}
\newcolumntype{Y}{>{\centering\arraybackslash}X}

\makeatletter\def\frontmatter@affiliationfont{\it\footnotesize}\makeatother

\newcommand*{\ttbar}{\ensuremath{t\bar{t}}\xspace}

\newcommand*{\pt}{\ensuremath{p_\mathrm{T}}\xspace}
\newcommand*{\spanet}{SPA-Net\xspace}

\begin{document}
  \title{Topological Reconstruction of Particle Physics Processes using Graph Neural Networks}
  \author{Lukas Ehrke$^a$}
  \email{lukas.ehrke@unige.ch}
  \affiliation{Département de physique nucléaire et corpusculaire, University of Geneva, Switzerland}
  \author{John Andrew Raine$^a$}
  \email{john.raine@unige.ch}
  \affiliation{Département de physique nucléaire et corpusculaire, University of Geneva, Switzerland}
  \author{Knut Zoch}
  \affiliation{Département de physique nucléaire et corpusculaire, University of Geneva, Switzerland}
  \affiliation{Laboratory for Particle Physics and Cosmology, Harvard University, Cambridge, 02138 MA, USA\\[1ex]}
  \author{Manuel Guth}
  \affiliation{Département de physique nucléaire et corpusculaire, University of Geneva, Switzerland}
  \author{Tobias Golling}
  \affiliation{Département de physique nucléaire et corpusculaire, University of Geneva, Switzerland}

  \begin{abstract}
    We present a new approach, the Topograph, which reconstructs underlying physics processes, including the intermediary particles, by leveraging underlying priors from the nature of particle physics decays and the flexibility of message passing graph neural networks.
    The Topograph not only solves the combinatoric assignment of observed final state objects, associating them to their original mother particles, but directly predicts the properties of intermediate particles in hard scatter processes and their subsequent decays.
    In comparison to standard combinatoric approaches or modern approaches using graph neural networks, which scale exponentially or quadratically, the complexity of Topographs scales linearly with the number of reconstructed objects. 
    
    We apply Topographs to top quark pair production in the all hadronic decay channel, where we outperform the standard approach and match the performance of the state-of-the-art machine learning technique.
  \end{abstract}
  
  \maketitle
  \renewcommand\thefootnote{\alph{footnote}}\footnotetext{These authors contributed equally to this work}\setcounter{footnote}{0}\renewcommand\thefootnote{\arabic{footnote}}

  \section{Introduction}
  
At the Large Hadron Collider (LHC) at CERN, beams of protons are collided together at incredibly high energies to probe the underlying nature of the universe.
From the objects recorded by the detectors on the LHC, the underlying physics processes in the collisions are attempted to be reconstructed.


In processes with a high multiplicity of objects, 
reconstructing intermediate particles, such as the two top quarks and $W$ bosons in top quark pair production, is a crucial and challenging component in analyses~\cite
{TOPQ-2012-08, CMS-TOP-14-022, ATLAS:2015pfy,  TOPQ-2015-03, TOPQ-2016-01,  CMS:2016oae, CMS-TOP-17-008, CMS:2018htd,  CMS:2018quc, ATLAS:2018fwq, ATLAS:2019guf, ATLAS:2019hxz, CMS:2021vhb, ATLAS:2022waa, TOPQ-2015-01, CMS:2015cal, TOPQ-2018-18}.
This is of particular interest when measuring the mass of an intermediary particle, or its kinematic properties as part of a cross section measurement.


In this work we introduce Topographs, a new machine learning approach for reconstructing the full hypothesised decay chain from observed objects. It leverages state-of-the-art machine learning~(ML) with underlying priors from particle physics, through the nature of particle decays, to reconstruct underlying physics processes including the intermediary particles.
The computational complexity of Topographs scales linearly with object multiplicity, whereas alternative methods scale quadratically~\cite{Shlomi_2021,fenton2021permutationless} or exponentially~\cite{Erdmann_2014,HIGG-2017-03}.


We apply Topographs to reconstruct the underlying processes in the production of top quarks pairs in the all hadronic decay channel.
However, the architecture can be applied to any particle physics process and is not limited to event level reconstruction.

  \section{Motivation}
  
In the case of top quark pair production, with each top quark decaying hadronically $t\rightarrow Wb \rightarrow qq^{\prime}b$, each quark is expected to initiate a shower in the detector which is reconstructed as a jet, resulting in six jets. The number of combinations to match the reconstructed jets to quarks from the top quark pair system is computationally intractable and grows exponentially with additional jets reconstructed in the final state, even when taking underlying symmetries into account.
Solving the combinatorics of this system is a key challenge in the measurement of the top quark and $t\bar{t}$ pair, and one which is often computationally limited.
The combinatorics can be restricted by looking at event topologies where the high momentum of the top quarks results in all three quarks being reconstructed in a single large radius jet, however this restricts the phase space to such topologies which represent only a small fraction of all events.
  \section{Current Approaches}
  In top quark physics kinematic event reconstruction forms a key part of many measurements. The $\chi^2$ method~\cite{TOPQ-2015-03} and the Kinematic Likelihood fitter (KLFitter)~\cite{Erdmann_2014} have been employed in a large number of analyses~\cite{TOPQ-2012-08, CMS-TOP-14-022, ATLAS:2015pfy,  TOPQ-2015-03, TOPQ-2016-01,  CMS:2016oae, CMS-TOP-17-008, CMS:2018htd,  CMS:2018quc, ATLAS:2018fwq, ATLAS:2019guf, ATLAS:2019hxz, CMS:2021vhb, ATLAS:2022waa}.
In both approaches, all combinatorics of jet matching to final state quarks and gluons (partons) in the $t\bar{t}$ final state are tested with kinematic constraints based on the masses of the reconstructed $W$ bosons and top quarks as minimisation criteria.
In the case of KLFitter these are used in conjunction with  with transfer functions and the particle decay widths. 

Although good performance can be achieved with such an approach, as the number of jets in an event increases, as well as the multiplicity of final state objects to be reconstructed, the number of combinations increases exponentially.
For example, in an all hadronic top quark pair event with 6 jets, there are 720 potential combinations. This is reduced to 90 by exploiting underlying symmetries. However for events with 7 jets the combinations increase to 630, and for 8 jets they increase further to 2520. 
Furthermore, as the exact values of the mass of the top quark and $W$ boson are used to test the likelihood of a combination, this leads to a biased estimator which focusses on assigning jets which together are closest to the hypothesised particles mass, rather than exploiting all the information about the pairs or triplets of objects.
It also assumes that in all cases the top quarks and $W$ bosons are on-shell. 

Building on the previous combinatoric approaches, simple approaches using machine learning (ML) have been developed. Instead of finding the most probable assignment using just the masses of intermediary particles, machine learning discriminants are used to identify correct assignments, exploiting more information from the event~\cite{HIGG-2017-03}.
Nonetheless, these approaches still suffer from the same problems as the KLFitter and $\chi^2$ methods, with each combination needing to be tested to identify the most likely.

Another approach which uses more information from the event is the Matrix Element Method~(MEM)~\cite{Fiedler_2010,CMS-SMP-13-004,TOPQ-2015-01,HIGG-2017-03}.
The MEM not only attempts to match objects to the final state objects in an event, but directly assesses the likelihood of observing an event given the matrix element for a process. This can be evaluated for each potential combination with the highest resulting probability chosen as the correct assignment.
However, it is extremely slow and computationally intensive.
To calculate the likelihood of an event, an integral over the whole phase space of possible final state particle momenta must be performed.
It is also reliant on a transfer function, which is used to convert the jets, charged leptons and missing transverse momentum recorded by the detector to the partons, charged leptons and neutrinos before any hadronisation and detector effects. As there is no accurate function to model this, it is at best an approximation optimised by hand.
Normalising flows present a solution to the computational challenge and approximate functions~\cite{Butter:2022vkj}, however do not yet address the combinatoric solving.

\subsection*{State of the art}

The state of the art machine learning approach uses attention transformers~\cite{SAJANet,fenton2021permutationless,shmakov2021spanet} to identify the indices of final state objects coming from intermediate particles. In this approach no graph structure is used and only the permutation invariant collection of objects are considered. 
The complexity of the approach can be reduced by taking into account the symmetries, as performed in Refs.~\cite{fenton2021permutationless,shmakov2021spanet} (SPA-Net), corresponding to removing potential solutions in the combinatoric approaches, which leads to an overall complexity of $\mathcal{O}\left(N^2\right)$. 

Graph Neural Networks~\cite{battaglia2018relational,wu2020comprehensive} are also employed in HEP to associate objects to a common origin, for example in secondary vertex reconstruction~\cite{Shlomi_2021} and could similarly be applied to combinatoric solving at the event level. These approaches have fully connected graphs with $N(N-1)$ edges.

In addition to their reduced computational complexity in comparison to traditional approaches, both attention and GNN approaches also demonstrate reduction in biases towards particle masses, as often seen in the combinatoric approaches.
However, in both GNNs and \spanet the target is to identify the two triplets of objects which correspond to the decay of each top quark, neglecting the structure of the decay, and the properties of the intermediary particles.

Other approaches employ physics inspired layers in order to assign parton labels~\cite{Badea:2022dzb} or try to predict the properties of intermediate particles directly~\cite{Qiu:2022xvr}.

  \section{The Topograph}

The use of GNNs in high energy physics applications~\cite{Qu:2019gqs,Moreno:2019bmu,shlomi2020graph,Bernreuther:2020vhm,Ju:2020tbo,Guo:2020vvt,Dreyer:2020brq,Hariri:2021clz,DeZoort:2021rbj,Atkinson:2021nlt,Thais:2022iok,Gong:2022lye,CMS-DP-2020-002,ATL-PHYS-PUB-2022-027} is a recent development which is gaining in popularity.
However, graphs have also long been used to describe underlying processes occurring in particle physics in the form of Feynman diagrams. 

\begin{figure}[h]
    \centering
    \begin{subfigure}[b]{0.45\textwidth}
        \centering
        \resizebox*{0.75\textwidth}{!}{%
        \includegraphics{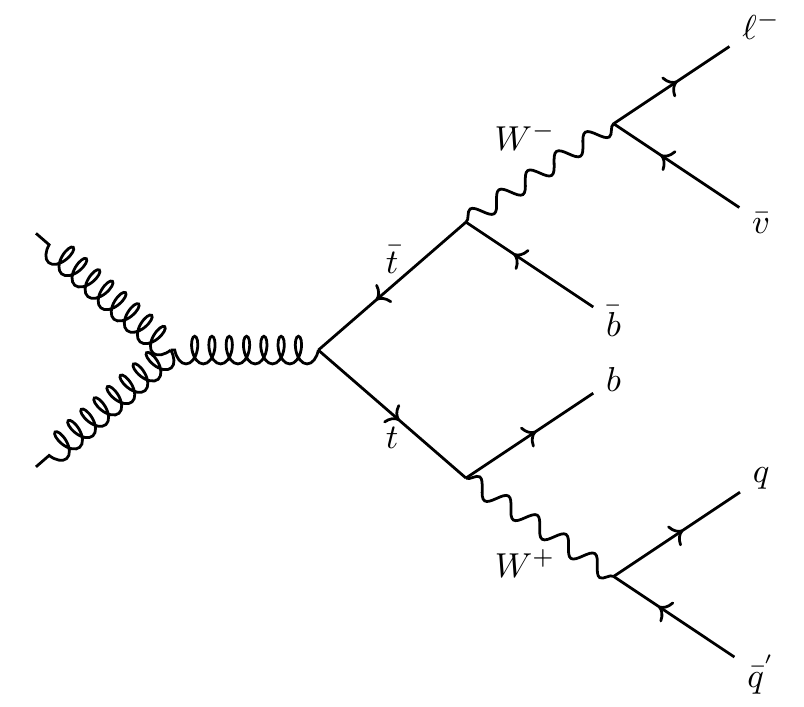}}
        \caption{Feynman diagram}
        \label{fig:topgraphscomp:ttbarfeyn}
    \end{subfigure}
    \hfill
    \begin{subfigure}[b]{0.45\textwidth}
        \centering
        \resizebox*{0.4\textwidth}{!}{%
        \includegraphics{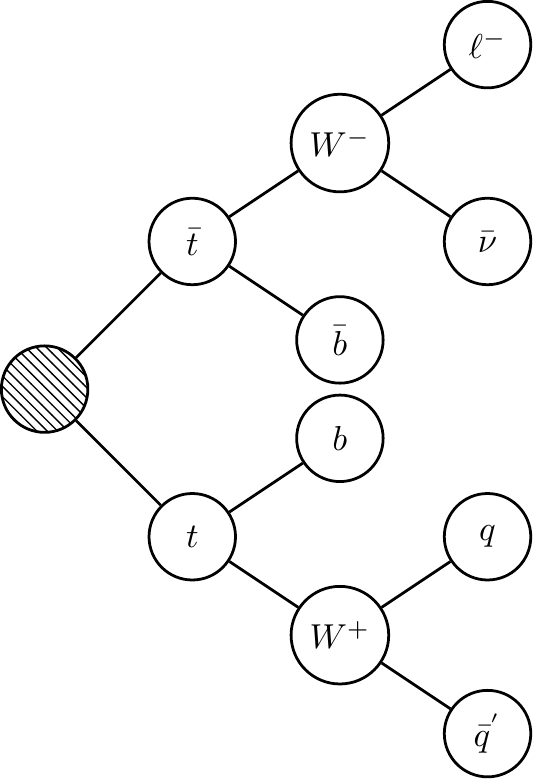}}
        \caption{Node and edge graph}
        \label{fig:topgraphscomp:ttbargraph}
    \end{subfigure}
    \caption{Two graph representations of top pair production, with one top decaying semi-leptonically and the other hadronically. The production mechanism in \subref{fig:topgraphscomp:ttbarfeyn} is shown to be gluon fusion however in \subref{fig:topgraphscomp:ttbargraph} it is represented by the hashed circle. Time flows from left to right in both graphs.}
    \label{fig:topgraphscomp}
\end{figure}

In a Feynman diagram, the vertices (nodes) represent the interactions between particles and the edges represent the particles themselves. This representation can equally be converted into a node-and-edge graph by representing the particles as nodes, and defining edges based on the vertices. An example of the Feynman digram and one such graph for top quark pair production is shown in \cref{fig:topgraphscomp}.
In current graph based approaches this physical inspired graph representation
is not exploited, and instead fully connected graphs are constructed from all objects recorded by the detector.

We introduce a new method, called the Topograph, which builds upon the structure seen in \cref{fig:topgraphscomp:ttbargraph} to define neural networks which can be used to predict the correct edges from final state particles and the properties of intermediate particles.

Firstly, the intermediate particles in a chosen physics process are injected into the graph. Secondly, instead of connecting all objects to one another, they are instead connected to all of their potential mother particles.
From this new graph, instead of identifying the edges based on a shared origin, true edges are identified as being between a mother and daughter particle, and the result is the reconstruction of the Feynman diagram as depicted in \cref{fig:topgraphscomp}.

Secondly, as the intermediate particles are now represented by their own nodes,  during training the kinematic properties of the injected nodes are predicted as auxiliary tasks with dedicated regression networks for each particle type.
Edges in GNNs are not only used to identify connections but also to propagate information. By utilising the edges of the Topograph as message passing layers to update the properties of the injected nodes, properties of the intermediary particles are extracted from the graph  and regression networks can predict their kinematic properties. 
This is advantageous both through improvements in the edge classification, but also from the additional extracted information.

\begin{figure}[h]
    \centering
    \begin{subfigure}{0.45\textwidth} 
        \centering
        \includegraphics{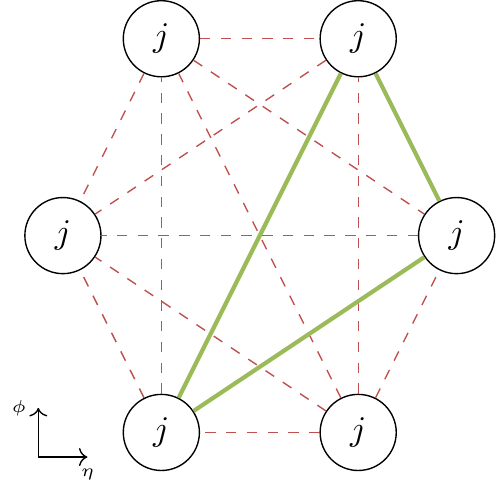}
        \caption{Fully connected graph}
        \label{fig:topedges:fcg}
    \end{subfigure}
    \begin{subfigure}{0.45\textwidth} 
        \centering
        \includegraphics{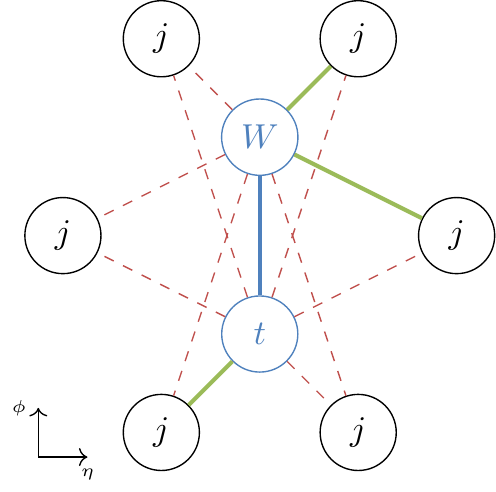}
          \caption{Topograph}
          \label{fig:topedges:topograph}
    \end{subfigure}
\caption{Comparison of the edges in \subref{fig:topedges:fcg} fully connected GNNs 
and \subref{fig:topedges:topograph} Topographs, when identifying the three jets $j$ which originate from a top quark decay.
The blue nodes in \subref{fig:topedges:topograph} represent the injected nodes, a key component of the Topograph, for both the top quark ($t$) and $W$ boson ($W$). The true edges, which identify jets which \subref{fig:topedges:fcg} originate from the same top quark or \subref{fig:topedges:topograph} reconstruct the decay chain, are green. All false edges in the graphs are dashed red lines. The predefined connection between $t$ and $W$ boson is blue.}
\label{fig:topedges}
\end{figure}

One leading advantage of Topographs over fully connected GNNs can be seen in Fig.~\ref{fig:topedges}, showing the simple case of identifying the jets from the decay of a single top quark.
In comparison to the fully connected GNN, which has $N\left(N-1\right)$ edges, the Topograph only requires $\mathcal{O}\left(N\right)$, which scales linearly with the number of intermediary particles; in this case there are $2N$ edges. 
For cases where $N>M+1$, where $M$ is the number of intermediary particles, there are always fewer edges associated to reconstructed objects in a Topograph than a fully connected GNN.

Furthermore, having a handle on the underlying kinematics of the intermediary particles with auxiliary tasks should improve the resolution and accuracy of current differential measurements, which are often a function of intermediate particle kinematics.
These properties could also be used in a likelihood test of the process, as is done in the MEM, instead of only using objects recorded by the detector.

With Topographs, complex underlying physics processes can be injected as priors by changing the injected particles and their potential connections. This enables additional information to be included when designing and training the networks over standard approaches.

\subsection*{Building blocks}

\begin{figure}[h]
    \centering
    \begin{subfigure}{0.5\textwidth} 
      \centering
      \includegraphics{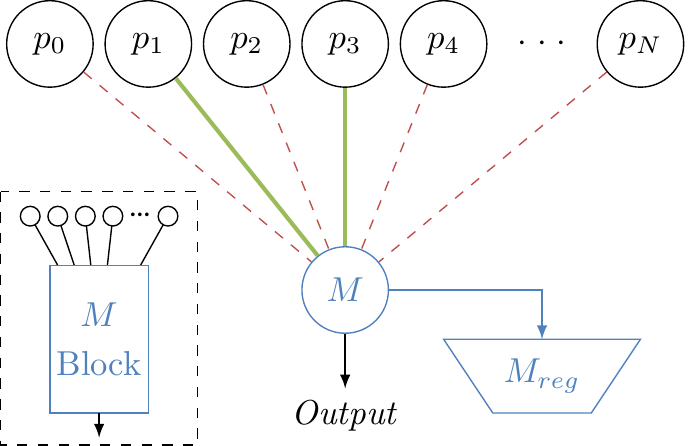}
    \end{subfigure}
    \begin{subfigure}{0.4\textwidth} 
      \centering
      \includegraphics{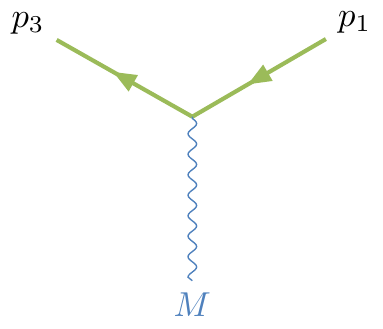}
    \end{subfigure}
    \caption{The particle block for a given mother particle $M$. 
    All input particles are connected to the mother particle by an edge; for illustrative purposes the true edges are shown in green and the false edges are represented by dashed red lines.
    The trapezoid $M_\mathit{reg}$ is the regression network which predicts the kinematic properties of $M$. 
    The inset is the notation used to represent the whole particle block.
    Shown alongside the Feynman diagram representing the process, with time running from bottom to top.}
    \label{fig:pblock:one}
\end{figure}

Although the Topograph could be visualised as one large graph with injected nodes and edges, we break them down into a simple set of building blocks.
The core building block of the Topograph is the particle block.
It includes the edge definitions between an input set of particles, or nodes, and the target mother particle. In addition, it contains the regression network used to predict the kinematic properties of the injected mother particle.
If a Topograph is visualised as a Feynman diagram of a process, a particle block is the subcomponent which determines the correct connections to recreate a single vertex alongside the properties of the incoming particle.
The basic representation of a particle block for a mother particle $M$ is depicted in \cref{fig:pblock:one} alongside a representative Feynman diagram vertex.
The injected mother particle $M$ can be initialised with random values, or using information extracted from all potential daughter particles.
Its properties are learned from message passing layers between itself and the input particles.

\begin{figure}[h]
  \centering
  \begin{subfigure}[b]{0.5\textwidth} 
    \centering
    \includegraphics{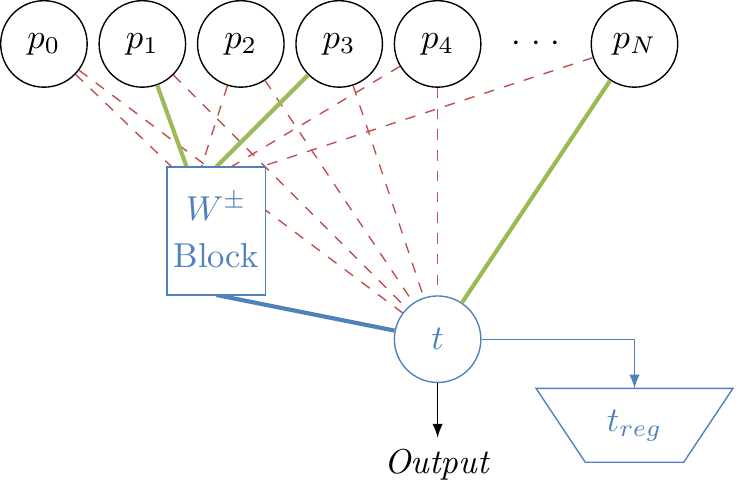}
  \end{subfigure}
  \begin{subfigure}[b]{0.4\textwidth} 
    \centering
  \includegraphics{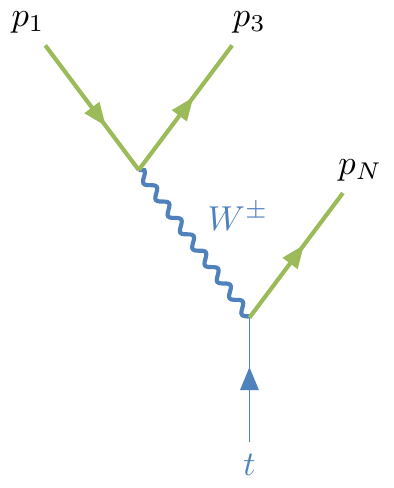}

  \end{subfigure}
  \caption{Particle block of a top quark $t$. A particle block for the $W$ is nested within the $t$ block. 
  A connection between the $W$ boson and the top quark is predefined (shown in blue).
  Shown alongside the corresponding Feynman diagram.}
  \label{fig:pblock:top}
\end{figure}

Any hypothesised process can be described by combining multiple particle blocks into a single network, connecting them to the input particles and to one another.
Edges between objects and particle blocks can also be predefined, for example between the particle blocks for a $W$ boson and a top quark, as shown in \cref{fig:pblock:top}.

\subsection*{Assembling a neural network with Topographs}

For many processes, there will be more than one reconstructed object type or final state particle in the chosen process. To address this a set of neural networks $\phi_p$, one for each particle type $p$, can be incorporated into the Topograph model as a series of particle embedding networks. 
Furthermore, in order to maximise initial information exchange before the Topograph it may be beneficial to include a normal message passing layer before the Topograph.
Options for this layer include attention transformers and standard message passing GNN layers.
Furthermore, in complex processes it is possible to define which edges are to be predicted and which are fixed. For example, two leptons in the production of a $Z$ boson in association with two top quarks could be set to originate from the $Z$ boson, or one from each $W$ boson.

  \section{Solving combinatorics in $\mathrm{t\bar{t}}$ events}
For an initial application and for direct comparison with other state-of-the-art methods we apply Topographs to top quark pair production with both tops decaying hadronically. 
We compare the performance of our method to a benchmark non-ML approach used in many top quark analyses, the $\chi^2$ method, and the state-of-the-art ML approach, \spanet~\cite{shmakov2021spanet}.
All models are trained and evaluated on the same dataset.

20 million $t\bar{t}$ events with a centre of mass energy $\sqrt{s}=13$~TeV are simulated using MadGraph5\_aMC@NLO~\cite{MadGraph}~(v3.1.0), with decays of top quarks and $W$~bosons modelled with MadSpin \cite{MadSpin}, with both $W$~bosons decaying to two quarks.%
\footnote{Dataset available at \url{https://zenodo.org/record/7737248} \cite{zoch_knut_2023_7737248}}
The parton shower and hadronisation is performed with Pythia~\cite{Pythia} (v8.243).
The detector response is simulated using Delphes~\cite{Delphes} (v3.4.2) with a parametrisation similar to the response of the ATLAS detector.
Jet clustering is performed using the anti-$k_{t}$ algorithm~\cite{AntiKt} with a radius parameter $R=0.4$ using the FastJet~\cite{FastJet} package.
Jets originating from $b$-quarks ($b$-jets) are identified with a simple binary discriminant corresponding to an inclusive 70\% signal efficiency.

For training, events with at least six jets are selected, keeping up to 16 jets per event as ordered by their transverse momentum.
Truth matching of jets to the partons in the hard scatter is performed using a cone of $\Delta R < 0.4$.
Events with partons matched to multiple jets, or jets matched to multiple partons are discarded.
Events are further required to have zero reconstructed leptons, though no requirement on the number of $b$-jets. 
Finally, events are required to be fully reconstructable, where jets are matched to all six partons of the \ttbar decay.
After these requirements there are 1,340,000 training events and 71,000 validation events.

An additional 298,000 events are reserved for evaluating the final performance.
These also contain events where not all partons have a jet matched to them.
In 76,000 of these events it is possible to associate a jet to all six partons.
After requiring at least 2 $b$-jets in the event, there are 147,000 events of which 44,000 are fully reconstructable.

As inputs for both ML models the four momentum $(p_x, p_y, p_z, E)$ together with a boolean flag showing whether the jet was $b$-tagged or not are used.
The three momenta are normalised subtracting the mean and dividing by the standard deviation. 
The energy is normalised in the same way after applying the logarithm.
No normalisation or transformation is applied to the $b$-tagging flag.
As both models represent the single jets as nodes, additional information per jet can easily be added for both models as additional node features.

\subsection{Topograph implementation}

\begin{figure}[ht]
    \centering
    \includegraphics{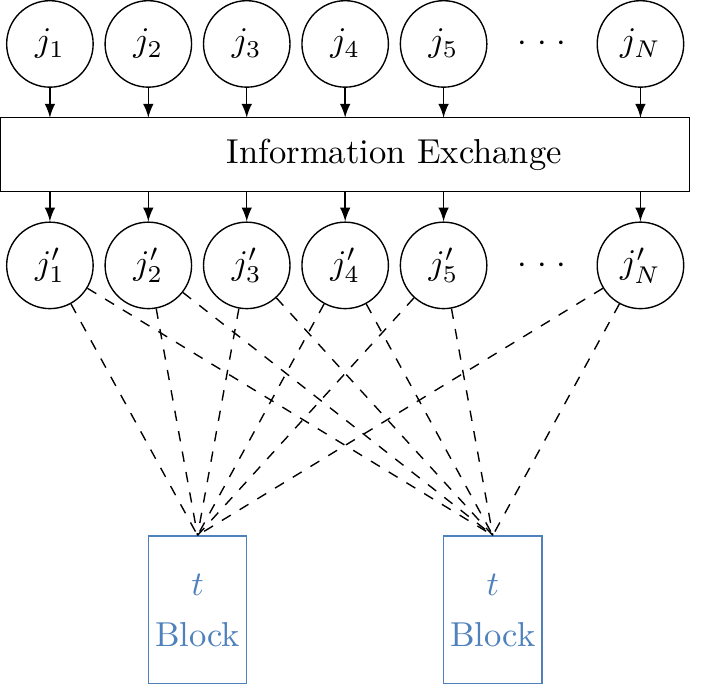}
    \caption{Topograph network for the $t\bar{t}$ process comprising two $t$ blocks.
    The jets are passed through an information exchange layer, using a fully connected graph message passible layer.
    All jets are connected to all possible mother particles, as shown by the dashed edges.
    The information exchange comprises multiple message passing graph layers and the node updates with the Topograph are performed $N$ times before the edge values are used for parton assignment and the properties for the top quarks and $W$ bosons are extracted.}
    \label{fig:topott}
\end{figure}

Before the particle blocks for the two top quarks, two fully connected message-passing graph layers are used to provide an information exchange between the jets in the event and update the jet features. 
From these updated jets, the $W$ nodes are initialised using attention weighted pooling.
Two different networks are used to obtain two sets of attention weights, one for each of the $W$ nodes.
The top nodes are initialised in the same way from the jets but with the corresponding $W$ node concatenated to the pooled jet information.
The regression targets of the $W$ and top nodes are the truth level properties of the particles.
In both cases the three momentum $(p_x, p_y, p_z)$ is chosen as the regression target.
In our dataset the truth mass is fixed to the Monte Carlo mass values. 
The network structure is shown in \cref{fig:topott}.

Within both the information exchange and Topograph blocks, message passing is bidirectional with a separate edge for each direction. Shared weights are used for calculating the attention pooling weights for each category of edge, defined by the sending and receiving particle type.
Edge features are formed by concatenating the features of the sending node and the receiving node.
Edge features are persistent between updates, and after the first iteration the current values are concatenated to the new features.

Four message passing steps in the Topograph update the jets, edges and injected $W$ and top nodes.
After the final message passing step, the properties of the $W$ bosons and top quarks and particle matching scores are extracted.
There are four matching scores for each jet, from which six jets are assigned to the six partons.
The scores are calculated by passing the edge properties of the jet to $W$ and top node through a classification network to determine if it is a true edge.

\subsubsection*{Loss function}

The loss function is calculated from both the edge classification task and the regression tasks.
Edges are classified as true if a jet originated from the $b$-quark, in the case of the top nodes, or one of the two quarks from the decay of the $W$ boson, for the $W$ nodes.
Here, binary cross-entropy is used for the loss function,
and it is weighted using the number of true and false edges across the dataset in order to improve convergence during training.
For each regression task, the mean absolute error (MAE) function is used between the predicted values of the node and the true values of the top quark or $W$ boson.

Since it is not possible for the network to determine which top node corresponds to the top quark or the anti-top quark, a symmetrised version of the losses is calculated in which the loss is calculated for both possible cases.
The loss is then defined as the minimum of the two.
Since the $W$ boson nodes are directly connected to the top nodes, the loss terms for the $W$ bosons are included in the loss calculation without need for additional symmetrisation.
The full loss is given by
\begin{align*}
    \mathcal{L}_\mathrm{symm} = \min(&\mathcal{L}(t_1, t) + \mathcal{L}(W_1, W^+) \\&+ \mathcal{L}(t_2, \bar{t}) + \mathcal{L}(W_2, W^-), \\
    &\mathcal{L}(t_1, \bar{t}) + \mathcal{L}(W_1, W^-) \\&+ \mathcal{L}(t_2, t) + \mathcal{L}(W_2, W^+)),
\end{align*}
where $\mathcal{L}(p_i, p)$ corresponds to the combined edge classification loss and regression loss for each of the injected particles $p_i \in t_1$, $t_2$, $W_1$ and $W_2$,
with respect to the truth particle $p \in t$, $\bar{t}$, $W^+$, $W^-$.



\subsubsection*{Parton assignment} 
Several options could be tested for assigning jets to the partons for the edge score.
For this initial study a simple iterative approach is chosen.
First, the edge with the highest score is labelled as a true edge, with the jet assigned to this parton.
Next all edges connected to the corresponding jet and parton are removed, and the next highest edge is chosen.
In the all hadronic \ttbar case, there is one parton per top quark, corresponding to the $b$-quarks in the decay, and two per $W$ boson.
This assignment procedure is repeated until all six partons are assigned, and results in a solution where neither a jet or parton can be assigned to multiple targets.


\subsection{Reference methods}

\subsubsection*{$\chi^2$ implementation}
The $\chi^2$ score for \ttbar decays used in this work is given by
\begin{align*}
    \chi^2 = \frac{(m_{b_1q_1q_2} - m_t)^2}{\sigma_t^2} + \frac{(m_{b_2q_3q_4} - m_t)^2}{\sigma_t^2}\\ + \frac{(m_{q_1q_2} - m_W)^2}{\sigma_W^2} + \frac{(m_{q_3q_4} - m_W)^2}{\sigma_W^2},
\end{align*}
where $m_{b_iq_jq_k}$ is the invariant mass of the jets in the permutation, $m_t$ and $m_W$ are the masses of the top quark and W boson, and $\sigma_t$ and $\sigma_W$ are the widths of the top quark and W boson in the dataset.
The values for $m_t$ and $m_W$ are obtained by taking the mean of the reconstructed invariant masses in our dataset, while $\sigma_t$ and $\sigma_W$ are set to the standard deviation.
The calculated values are $m_t=169.8\ \mathrm{GeV}$, $m_W=81.0\ \mathrm{GeV}$, $\sigma_t=29.0\ \mathrm{GeV}$, and $\sigma_W=18.5\ \mathrm{GeV}$.
Jets associated to the $b$-quark of the top decays are required to be $b$-tagged, while no requirement is placed on the jets associated to the $W$ boson decays.

\subsubsection*{\spanet implementation}
The implementation is taken from the github repository\footnote{\url{https://github.com/Alexanders101/SPANet/tree/v1.0}}.
The hyperparameters of the model are optimised for the dataset using fully matched events, with the values in Ref.~\cite{fenton2021permutationless} using \emph{v1.0} resulting in the highest overall efficiencies.
All \spanet models are trained for 100 epochs, taking the weights after the epoch with the lowest loss on the validation set.

\subsection{Partial event trainings}

It is also possible to train Topographs on events in which not all partons from the $t\bar{t}$ decay are matched to jets.
For these events, the same network is used but the edge classification and regression loss terms are not considered for the $W$ boson or top quarks which have partons not able to be matched to jets. 
In the case of the $b$-quark from the top quark decay not being matched to a jet, the $\mathcal{L}(W_i,W^+)$ term is still considered.
Where a quark from the $W$ boson decay is missing, both the top quark and $W$ boson loss terms are not considered.
At least one $W$ boson is required to be fully reconstructable, with both quarks matched to jets in the event.

As introduced in Ref.~\cite{shmakov2021spanet}, \spanet can also be trained on non-fully reconstructable events, however in comparison to Topographs this is only at the level of each top quark.
When a $b$-quark from the top quark is missing, the corresponding $W$ boson is also not considered.
For a fairer comparison with both models trained on the same number of events, we compare models trained only on fully reconstructable events.
Results for the Topograph and \spanet models trained with partial events are presented in \cref{app:partialspa}.

  
  \section{Results}
  
\begin{table*}[tb]
    \centering
    \caption{Event reconstruction efficiencies (\%) for the $\chi^2$ method, the \spanet model and our Topograph model in different jet and $b$-jet multiplicities.
    The mean and standard deviation are calculated using five trainings, and the best model corresponds to the training with the highest efficiency for events with at least six jets and exactly two $b$-jets.
    The highest efficiency is highlighted in bold.}
    \label{tab:efficiencies}
    \begin{tabularx}{0.95\textwidth}{Y|Y|YY|YYY} 
        \toprule
         & & \multicolumn{2}{c|}{Mean} & \multicolumn{3}{c}{Best} \\ 
         $N_{jets}$ & $N_{b-\mathrm{jets}}$ & \spanet & Topograph & \spanet & Topograph & $\chi^2$ \\ 
         \midrule
         \hphantom{$\geq$}$6$ & \hphantom{$\geq$}$2$ & $81.43\pm0.03$ & $\mathbf{81.54\pm0.12}$ & $81.47\pm0.31$ & $\mathbf{81.70\pm0.31}$ & $72.73\pm0.36$ \\
         \hphantom{$\geq$}$6$ & $\geq2$ & $79.45\pm0.07$ & $\mathbf{79.76\pm0.14}$ & $79.56\pm0.30$ & $\mathbf{79.95\pm0.29}$ & $70.94\pm0.33$ \\
         \hphantom{$\geq$}$7$ & \hphantom{$\geq$}$2$ & $64.47\pm0.08$ & $\mathbf{65.46\pm0.20}$ & $64.69\pm0.44$ & $\mathbf{65.90\pm0.44}$ & $54.28\pm0.46$ \\
         \hphantom{$\geq$}$7$ & $\geq2$ & $62.38\pm0.12$ & $\mathbf{63.51\pm0.25}$ & $62.86\pm0.40$ & $\mathbf{63.94\pm0.40}$ & $52.11\pm0.41$ \\
         $\geq6$ & \hphantom{$\geq$}$2$ & $68.48\pm0.06$ & $\mathbf{69.12\pm0.19}$ & $68.65\pm0.25$ & $\mathbf{69.54\pm0.25}$ & $58.57\pm0.26$ \\
         $\geq6$ & $\geq2$ & $65.76\pm0.08$ & $\mathbf{66.59\pm0.22}$ & $66.05\pm0.23$ & $\mathbf{67.03\pm0.22}$ & $55.90\pm0.24$ \\
         \bottomrule
    \end{tabularx}
\end{table*}
  For the evaluation of the performance only events with at least 2 $b$-tagged jets are considered. This is a common requirement in physics analysis to reduce the contribution from the multi-jet background.
This requirement is not imposed during training as it is found to reduce the performance as a result of the reduction in training statistics.

\subsection{Jet parton assignment}

Although Topographs predict the kinematics of injected particles,
the most important measure of performance is the association of jets to partons.
Efficiencies of reconstructing the whole event correctly are shown in \cref{tab:efficiencies} for varying jet and $b$-jet multiplicities.
Only events which are fully reconstructable are considered.
Topographs and \spanet show very similar performance across all jet multiplicities, with sub-percent differences in the reconstruction efficiencies.
Both clearly outperform the $\chi^2$ method by around $10\%$ with larger differences at higher jet multiplicities.

\begin{figure}[ht]
    \centering
    \includegraphics[width=0.45\textwidth]{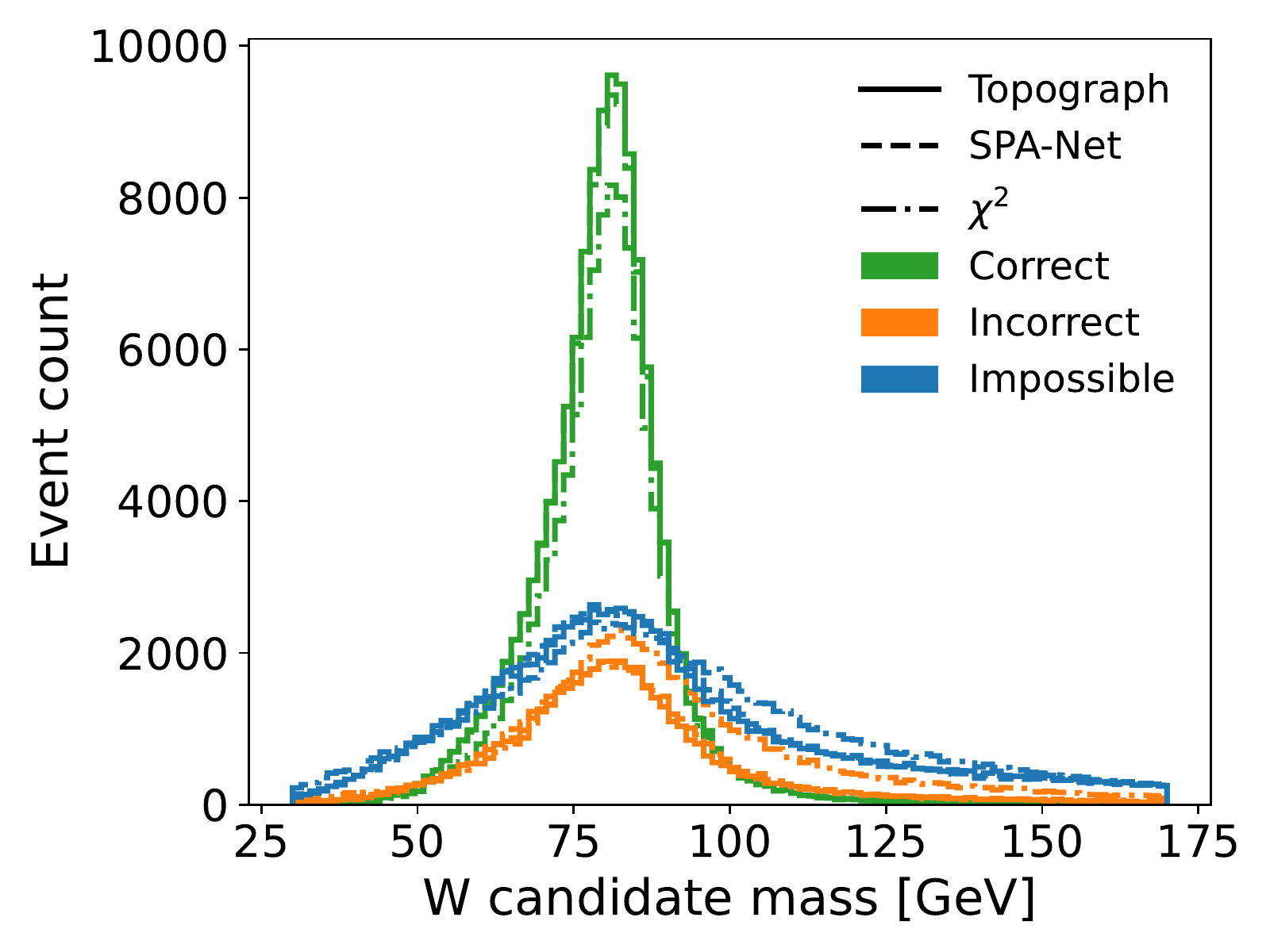}
    \caption{Reconstructed $W$ boson invariant mass $m_W$ using jets assigned by the Topograph (solid), \spanet (dashed), and the $\chi^2$ method (dash-dot).
    Events are categorised into correct (green) and incorrect assignments (orange), with events missing a parton at reconstruction level labelled as impossible (blue).}
    \label{fig:reco_mass_comparison_w}
\end{figure}

\begin{figure}[ht]
    \centering
    \includegraphics[width=0.45\textwidth]{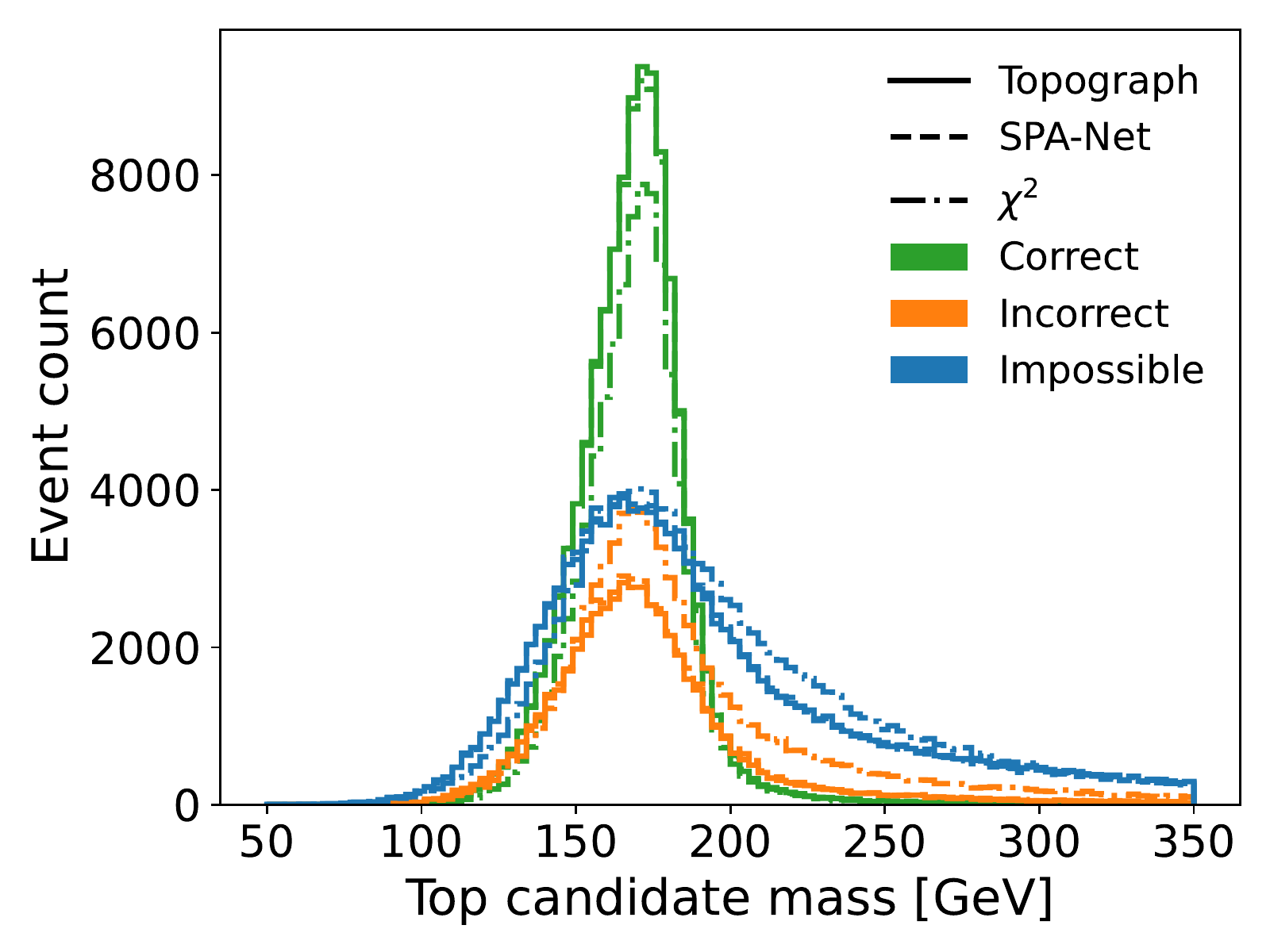}
    \caption{Reconstructed top quark invariant mass $m_{top}$ using jets assigned by the Topograph (solid), \spanet (dashed), and the $\chi^2$ method (dash-dot).
    Events are categorised into correct (green) and incorrect assignments (orange), with events missing a parton at reconstruction level labelled as impossible (blue).
    }
    \label{fig:reco_mass_comparison_top}
\end{figure}

\begin{table*}[tb]
    \centering
    \caption{Percentage of events with at most $N$ incorrectly matched partons using the $\chi^2$ method, \spanet and Topographs.
    Events are categorised based on the number of partons matched to jets at truth level.
    The highest efficiency is highlighted in bold.}
    \label{tab:correct_jets}
    \begin{tabularx}{0.95\textwidth}{YY|YYYYYY}
        \toprule
        $N_{partons}$ &  & \multicolumn{6}{c}{Incorrectly matched partons [\%]} \\
        matched & Model & 0 & $\leq1$ & $\leq2$ & $\leq3$ & $\leq4$ & $\leq5$  \\
         \midrule
        \multirow{3}{*}{3} & \spanet & $34.34\pm0.65$ & $74.88\pm0.59$ & $98.03\pm0.19$ & $100.00\pm0.00$ & - & -\\
         & Topograph & $\mathbf{36.57\pm0.66}$ & $77.80\pm0.57$ & $98.70\pm0.15$ & $100.00\pm0.00$ & - & -\\
         & $\chi^2$ & $34.30\pm0.65$ & $\mathbf{80.12\pm0.54}$ & $\mathbf{99.05\pm0.13}$ & $100.00\pm0.00$ & - & -\\
         \midrule
        \multirow{3}{*}{4} & \spanet & $35.69\pm0.28$ & $64.53\pm0.28$ & $90.99\pm0.17$ & $99.41\pm0.04$ & $100.00\pm0.00$ & - \\
         & Topograph & $\mathbf{37.74\pm0.28}$ & $\mathbf{67.06\pm0.27}$ & $\mathbf{92.21\pm0.16}$ & $99.53\pm0.04$ & $100.00\pm0.00$ & - \\
         & $\chi^2$ & $31.01\pm0.27$ & $63.84\pm0.28$ & $92.19\pm0.16$ & $\mathbf{99.73\pm0.03}$ & $100.00\pm0.00$ & - \\
         \midrule
        \multirow{3}{*}{5} & \spanet & $44.74\pm0.19$ & $64.87\pm0.18$ & $86.92\pm0.13$ & $97.91\pm0.05$ & $99.94\pm0.01$ & $100.00\pm0.00$ \\
         & Topograph & $\mathbf{46.57\pm0.19}$ & $\mathbf{66.88\pm0.18}$ & $\mathbf{88.58\pm0.12}$ & $\mathbf{98.45\pm0.05}$ & $\mathbf{99.96\pm0.01}$ & $100.00\pm0.00$ \\
         & $\chi^2$ & $32.36\pm0.18$ & $58.27\pm0.19$ & $84.70\pm0.14$ & $98.21\pm0.05$ & $99.94\pm0.01$ & $100.00\pm0.00$ \\
         \midrule
        \multirow{3}{*}{6} & \spanet & $66.05\pm0.23$ & $74.43\pm0.21$ & $90.06\pm0.14$ & $96.22\pm0.09$ & $99.52\pm0.03$ & \hphantom{$1$}$99.98\pm0.01$ \\
         & Topograph & $\mathbf{67.03\pm0.22}$ & $\mathbf{75.71\pm0.20}$ & $\mathbf{91.33\pm0.13}$ & $\mathbf{97.03\pm0.08}$ & $\mathbf{99.70\pm0.03}$ & $\mathbf{100.00\pm0.00}$ \\
         & $\chi^2$ & $55.90\pm0.24$ & $64.93\pm0.23$ & $84.14\pm0.17$ & $93.97\pm0.11$ & $99.43\pm0.04$ & \hphantom{$1$}$99.98\pm0.01$ \\
        \bottomrule
    \end{tabularx}
\end{table*}

\Cref{fig:reco_mass_comparison_w,fig:reco_mass_comparison_top} show the reconstructed $W$ boson and top quark invariant masses.
The distributions are shown separately for events with the correct and incorrect jet assignments, as well as events which are not possible to be fully reconstructed due to not all partons being matched to reconstructed jets (impossible).
Events where the correct triplet is assigned to the top quark but with the incorrect matching account for 5\% of the incorrect events for Topographs and \spanet, and 3\% for $\chi^2$.
All methods are able to reconstruct candidates closest to the particle mass, with a broader distribution for events without the corresponding partons matched to jets in the event.
Normalised and stacked distributions can be found in \cref{app:plots}.
\spanet and Topographs have very similar distributions showing no difference in potential sculpting of the mass distributions.
The $\chi^2$ method has distributions which are shifted towards slightly higher values compare to \spanet and Topographs.

For events which do not contain all partons in the final state, it is of interest to see how many of the partons are correctly matched to jets.
In \cref{tab:correct_jets} the matching efficiencies of only the partons which are present in the event are compared for the three approaches.
The Topograph performs slightly better than \spanet, with higher efficiencies for correctly matching all available partons or only one incorrectly matched parton.
Both Topographs and \spanet are substantially better than the $\chi^2$ at correctly identifying all available partons.
It should be noted that the perfect reconstruction in this case is substantially lower than the fully reconstructable events in \cref{tab:efficiencies}.

In \cref{tab:flavour_jet_efficiencies} we break down the matching efficiencies of jets in incorrect events by the truth flavour of the jet. 
The jet truth flavour is assigned using $\Delta R$ matching to $B$- and $C$-hadrons.
We observe that Topographs and \spanet have similar jet matching efficiencies for all jet flavours.
Topographs have a slightly higher efficiency for $b$-jets, which correspond to the jets coming from the $b$-quarks in the top quark decays as well as in matching light jets to the quarks in the $W$ boson decay.
The $\chi^2$ method has the largest efficiency drop for truth flavour $c$-jets.
This results from $c$-jets being mis-identified by the $b$-tagging, which in the $\chi^2$ method allows them to be matched to the $b$-quarks from the top decays.

\begin{table}[h]
    \centering
    \caption{Efficiencies (\%) of correctly associating a jet to its corresponding parton for the $\chi^2$ method, \spanet and Topographs.
    Jets are categorised into their truth flavour using parton matching. Only jets coming from the $\ttbar$ decays in events with at least one incorrect assignment are considered.
    The highest efficiency is highlighted in bold.}
    \label{tab:flavour_jet_efficiencies}    
    \begin{tabularx}{0.475\textwidth}{Y|YYY}
        \toprule
        Correct & \spanet & Topograph  & $\chi^2$ \\
        \midrule
        $b$-jets  & $58.94\pm0.29$ & $\mathbf{59.99\pm0.29}$  & $57.68\pm0.25$ \\
        $c-$jets  & $55.08\pm0.39$ & $\mathbf{55.81\pm0.39}$  & $50.38\pm0.34$ \\
        light jets  & $70.42\pm0.22$ & $\mathbf{72.01\pm0.22}$  & $68.51\pm0.20$ \\
        \bottomrule
    \end{tabularx}
\end{table}

\begin{table*}[tb]
    \centering
    \caption{Efficiencies (\%) of correctly associating a jet to the corresponding parton for events with at least one incorrect assignment for the $\chi^2$ method, \spanet and Topographs.
    Top quarks are classified by the momentum of the corresponding $b$-quark.
    The quarks from the $W$ boson decays are ordered by their \pt.
    The highest efficiency is highlighted in bold.
    }
    \label{tab:pt_jet_efficiencies}   
    \begin{tabularx}{0.95\textwidth}{YY|YYY}
        \toprule
         & Correctly identified & \spanet & Topograph & $\chi^2$ \\
        \midrule
        \multirow{3}{*}{\shortstack{top quark\\ (leading $b$)}} & $b$-quark  & $61.83\pm0.40$ & $\mathbf{62.94\pm0.40}$ & $59.17\pm0.35$ \\
        & Leading quark ($W$)  & $67.79\pm0.38$ & $\mathbf{68.92\pm0.39}$ & $61.80\pm0.35$ \\
        & Subleading quark ($W$)  & $63.22\pm0.40$ & $\mathbf{64.70\pm0.40}$ & $65.09\pm0.34$ \\
        \midrule
        \multirow{3}{*}{\shortstack{top quark\\ (subleading $b$)}} & $b$-quark  & $56.05\pm0.41$ & $\mathbf{57.03\pm0.41}$ & $56.20\pm0.36$ \\
        & Leading quark ($W$)  & $67.49\pm0.38$ & $\mathbf{68.22\pm0.39}$ & $61.30\pm0.35$ \\
        & Subleading quark ($W$)  & $66.41\pm0.39$ & $\mathbf{68.31\pm0.39}$ & $65.94\pm0.34$ \\
        \bottomrule
    \end{tabularx}
\end{table*}

The matching efficiencies of partons in incorrect events are shown in \cref{tab:pt_jet_efficiencies}.
Here we see that the $\chi^2$ method mostly performs worse for matching jets to the quarks in the $W$ boson decay.
Topographs and \spanet show similar behaviour, with the slightly better performance in Topographs for incorrect events evenly distributed across all partons.


\subsection{Interpreting edge scores}

Due to the individual edge scores, the confidence of the combinatoric assignment from Topographs can be obtained by aggregating the edge scores.
This could be used to filter events as likely to be incorrect, as well as identify events for which it is not possible to match jets to all partons.

\begin{figure}[h]
    \centering
    \includegraphics[width=0.45\textwidth]{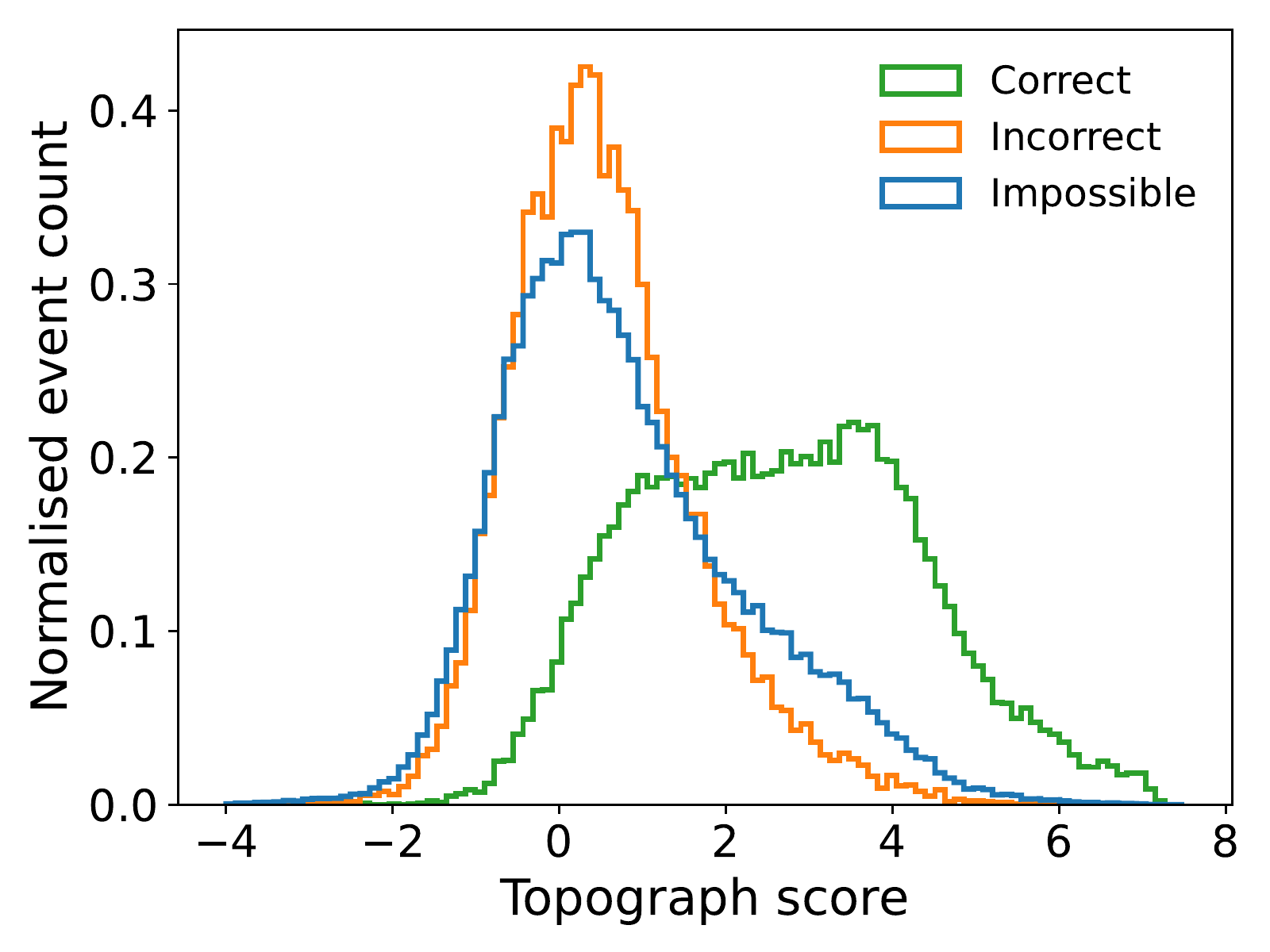}
    \caption{Event matching score for events with the correct assignment (green), incorrect assignment (orange) and non fully-reconstructable events (impossible, blue).}
    \label{fig:score_topograph_all}
\end{figure}

\begin{figure}[h]
    \centering
    \includegraphics[width=0.45\textwidth]{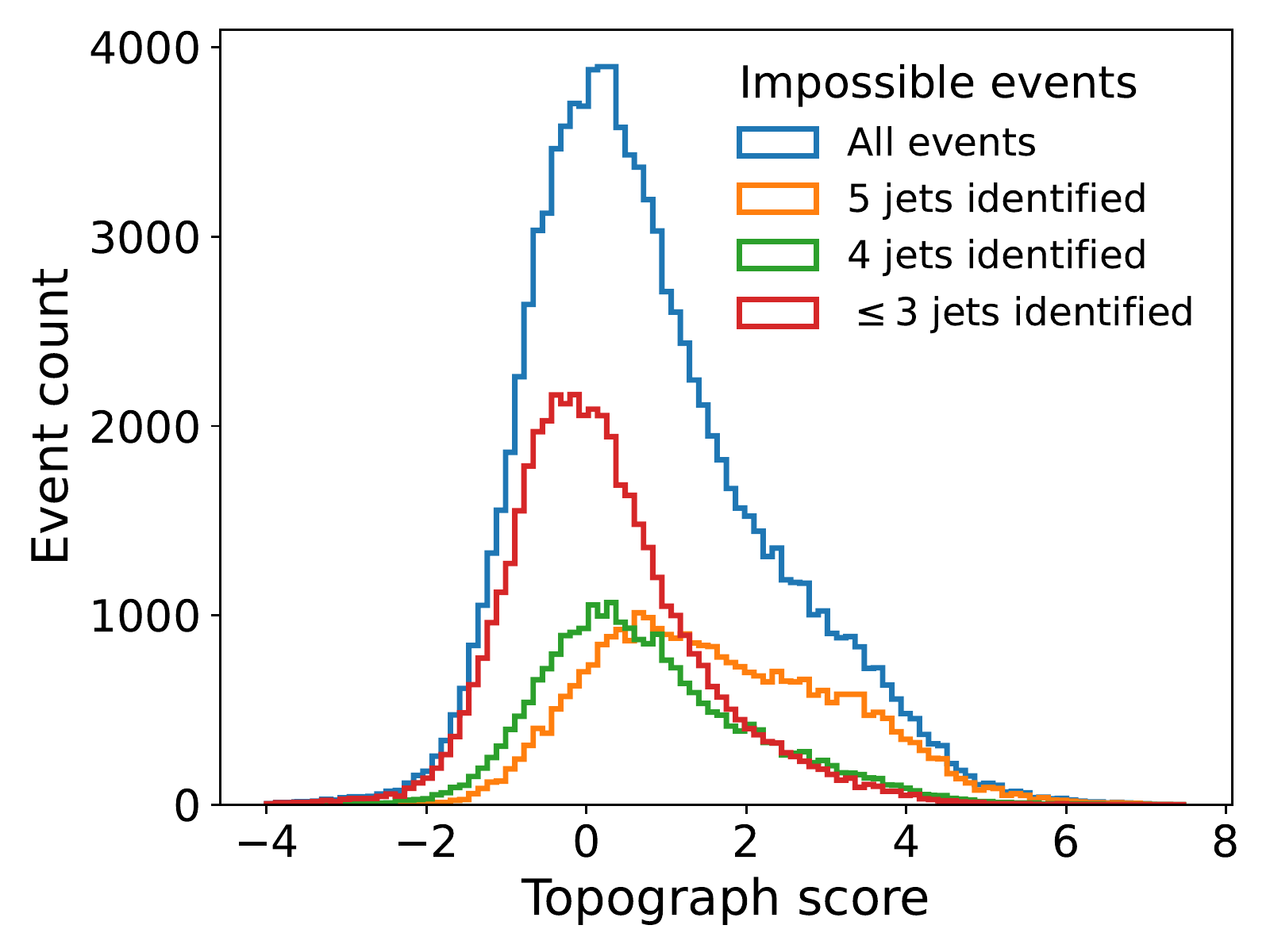}
    \caption{Event matching score for non-fully reconstructable events, categorised by the number of correctly assigned jets.}
    \label{fig:score_topograph_impossible}
\end{figure}

In \cref{fig:score_topograph_all} the product of assigned jet edge scores is used as an event matching score. 
There is reasonable separation between events with a correct assignment and those with the incorrect and impossible assignments.
However, a shoulder towards higher values is visible for the impossible events.

In \cref{fig:score_topograph_impossible} the impossible events are categorised into the number of jets which are correctly assigned.
It can be seen that the score is still high for events where there is a single parton missing, but all remaining partons are correctly matched to jets.
Whether using this score to remove events will benefit other down-stream applications, such as top quark mass measurements would need to be tested.

\subsection{Regression performance}

\begin{figure}[h]
    \centering
        \includegraphics[width=0.45\textwidth]{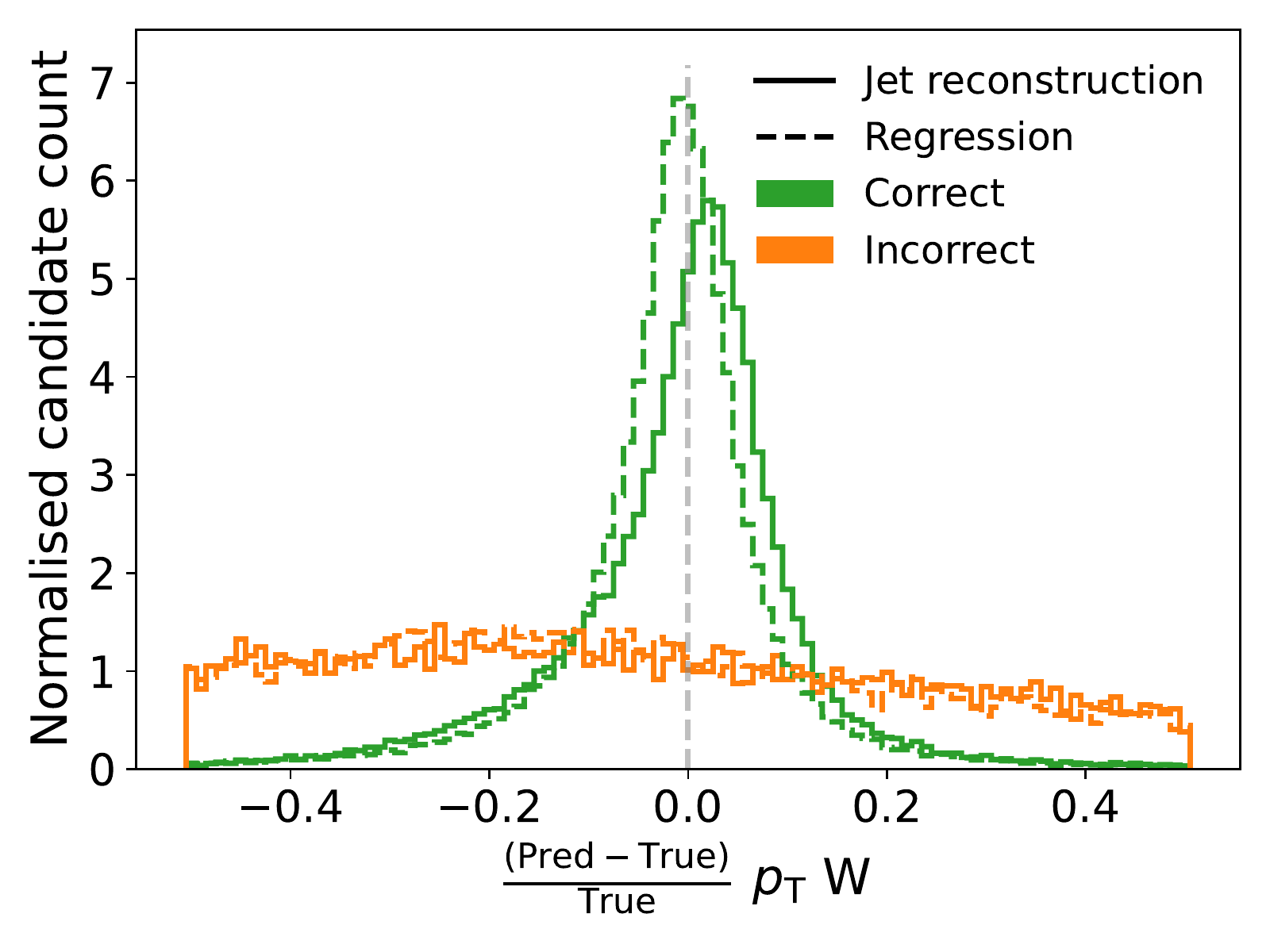}
    \caption{Resolution of the reconstructed \pt of the $W$ boson from the Topograph.
    Comparing the prediction from the invariant system of the assigned jets (solid line) and the Topograph regression network (dashed line) for correct assigned events (green) and incorrect assigned events (orange).}
    \label{fig:regression_pt_w}
\end{figure}

\begin{figure}[h]
    \centering
        \includegraphics[width=0.45\textwidth]{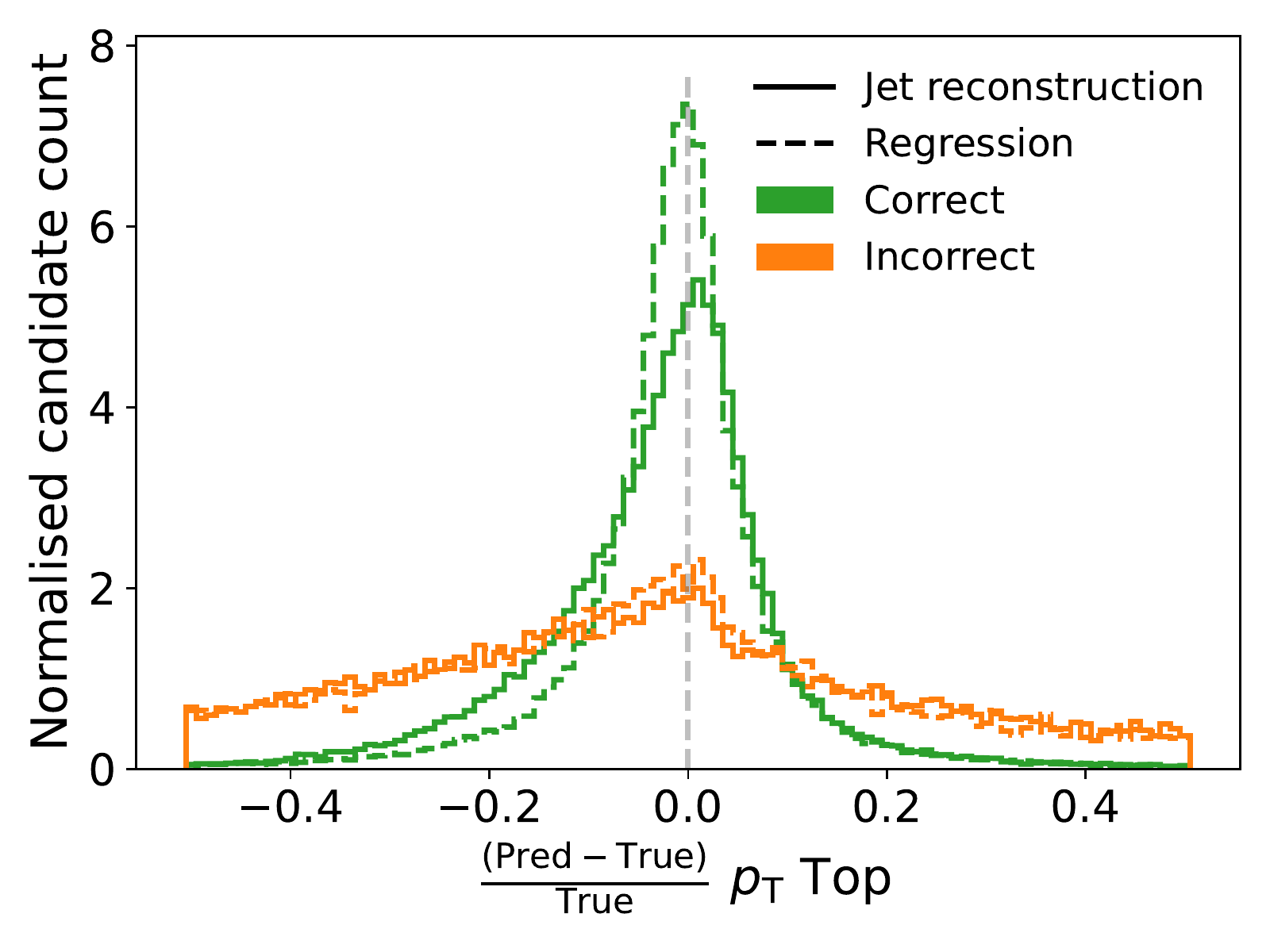}
    \caption{Resolution of the reconstructed \pt of the top quark from the Topograph.
    Comparing the prediction from the invariant system of the assigned jets (solid line) and the Topograph regression network (dashed line) for correct assigned events (green) and incorrect assigned events (orange).}
    \label{fig:regression_pt_top}
\end{figure}

Instead of reconstructing the top quark and $W$ boson kinematics from the matched jets, Topographs are trained to predict their properties directly.
\Cref{fig:regression_pt_w} compares the resolution of the $W$ boson \pt and \cref{fig:regression_pt_top} the top quark \pt from the Topograph regression networks and the invariant system of the assigned jets.
The predictions are shown for both correct and incorrect jet assignments.

For events with all jets correctly assigned, using the \pt values from the Topograph regression networks leads to a narrower peak than the reconstructed quantities, demonstrating a more accurate prediction.
They also show a reduced bias compared to the values reconstructed solely using the jets.
For events with incorrectly assigned jets, no difference is observed between the two predictions.
This shows that the Topograph is not learning to guess the \pt from the training data, but instead is using the associated jets.
Accurate reconstruction of the intermediate particles has a strong dependence on whether the model has correctly assigned jets to partons.
There are no biases observed using either method for the top quark and $W$ boson, and 
There is very little difference for reconstruction of the top quark $\eta$ and $\phi$ coordinates.
However reconstructing the $W$ boson $\eta$ and $\phi$ coordinates using the assigned jets performs better than the prediction from the network.
In all cases, the Topograph is trained to predict the momentum components $p_x, p_y, p_z$ rather than $p_\mathrm{T},\eta,\phi$.

  
  \section{Conclusion}
  In this work we have introduced Topographs, a novel approach for solving the combinatorics and reconstructing the topology of a particle physics process from final state objects reconstructed by a detector.
The performance matches the current state-of-the-art technique using symmetry preserving attention transformers, and surpasses the standard approach commonly used in analyses, with a computational complexity which scales only linearly with increasing final state object multiplicity.

The edge scores from Topographs can be combined into discriminants to assign a confidence to the jet-parton assignments, which could be useful in downstream applications.
Furthermore, the additional regression tasks included in Topographs demonstrate good predictive power with similar accuracy but reduced bias compared to using only the jets assigned to intermediate particles.
However, in both cases there remains room for improvement.

There are several other areas open for further optimisation.
Due to the message passing layers used to define the Topograph, it was found that fully connected graph layers between all jets for information exchange lead to faster convergence whilst training, and also resulted in requiring fewer learnable parameters in the model.
This causes the complexity of the network in this work to scale quadratically with the number of jets, and not linearly as can be achieved by only using the particle blocks in Topographs.
By moving to other architectures such as Transformer Encoders with cross-attention, the need for information exchange layers could be mitigated.
Alternative approaches for assigning jets to partons based on the edge scores could also improve the overall performance.

Applications of Topographs are not limited to the case study presented in this paper, and due to their modular nature Topographs can be generalised to almost all particle physics processes.
Their applications are also not limited to matching final state objects to an underlying physics process, but could also find use in jet identification, reconstructing displaced vertices from heavy flavour hadron decays or using the constituents in large radius jets in boosted topologies.

  \section*{Acknowledgements}
  The authors would like to acknowledge funding through the SNSF Sinergia grant called "Robust Deep Density Models for High-Energy Particle Physics and Solar Flare Analysis (RODEM)" with funding number CRSII$5\_193716$, the SNSF project grant 200020\_212127 called "At the two upgrade frontiers: machine learning and the ITk Pixel detector", and the Alexander von Humboldt foundation Feodor Lynen fellowship programme.

  \bibliography{bib/ATLAS.bib, bib/CMS.bib, bib/PubNotes.bib, bib/myrefs.bib}

  \appendix
  \section{Appendix}
  \subsection{Hyperparameters}

\Cref{tab:app_hyperparameters} shows the hyper parameters used for the training of the Topograph models presented in this paper.
The models were trained using \textsc{Tensorflow} v2.10~\cite{tensorflow2015-whitepaper}.

\begin{table}[htbp]
    \caption{Hyper parameters used for the training of Topographs}
    \label{tab:app_hyperparameters}
    \begin{tabular}{cc}
        \toprule
        Hyper parameter & Value \\
        \midrule
        Optimizer & \textit{AdamW} \\
        Epochs & 100 \\
        n original message passing & 2 \\
        n topograph updates & 4 \\
        lr schedule & cosine annealing \\
        initial learning rate & 0.001 \\
        decay steps & 2 epochs \\
        batch size & 256 \\
        pooling & attention \\
        W initialisation & attention pooling \\
        Top initialisation & attention pooling \\
        attention units & [32, 32, 1] \\
        activation & gelu \\
        normalisation & layer norm \\
        regression units & [64, 64, 3] \\
        edge classification units & [128, 128, 128, 1] \\
        graph processing units & [256, 256, 64] \\
        persistent edges & true \\
        classification loss & weighted binary crossentropy \\
        regression loss & mean absolute error \\
        \bottomrule
    \end{tabular}
\end{table}

\subsection{Example Topograph}

\begin{figure}[htbp]
    \centering
    \resizebox*{0.4\textwidth}{!}{
    \includegraphics{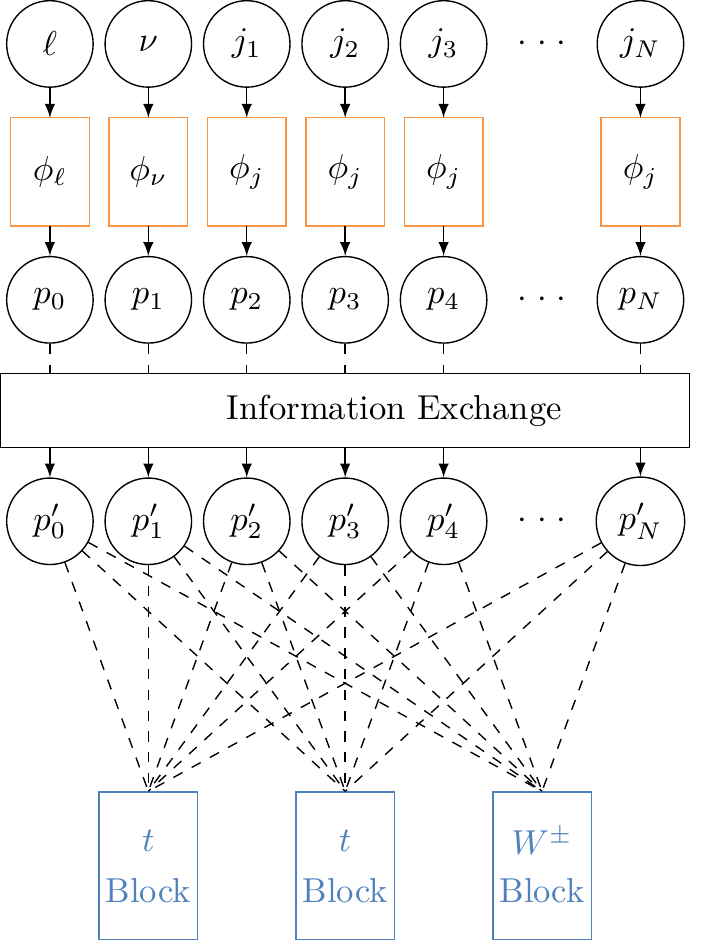}%
    }
    \caption{Topograph network for the $t\bar{t}W$ process comprising two $t$ blocks and a $W$ block.
    The input particles are first passed through an embedding network $\phi$ unique to each type of particle; either a jet $j$, lepton $\ell$ or neutrino $\nu$.
    Embedded particles are then passed through an (optional) information exchange layer.
    All particles are connected to all possible mother particles, as shown by the dashed edges.}
    \label{fig:topottW:embed-noprior}
\end{figure}

\begin{figure}[htbp]
    \centering
    \resizebox*{0.4\textwidth}{!}{
    \includegraphics{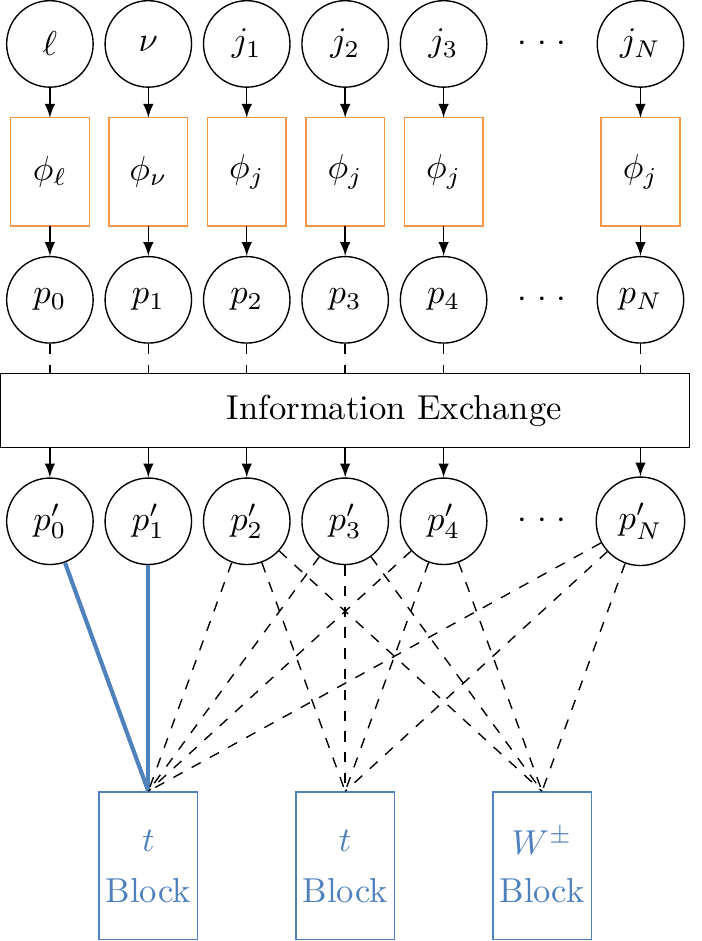}%
    }
  \caption{Topograph network for the $t\bar{t}W$ process comprising two $t$ blocks and a $W$ block
  where the $\ell$ and $\nu$ are predefined to connect to the $W$ in one top block.}
  \label{fig:topottW:embed-prior}

\end{figure}

A complete Topograph network is shown in \cref{fig:topottW:embed-noprior} for reconstruction of $t\bar{t}W$ in events containing exactly one lepton, one neutrino and multiple jets.
For the production of a top quark pair in association with a $W$ boson ($t\bar{t}W$), two top quark blocks and one $W$ block are connected to the input particles, but not to one another. The Topograph network is trained to identify the edges of the true daughter particles of each particle in the process, and predicts the kinematics of the two top quarks and the three $W$ bosons.
As Topographs are defined to represent an underlying physics process, it is also a choice whether to predefine whether the lepton and neutrino originate from a $W$ boson coming from a top decay, or the additional $W$ boson.
\Cref{fig:topottW:embed-prior} shows a Topograph where the lepton is required to come from a top decay.

\subsection{Comparisons with partial event trainings}
\label{app:partialspa}

Instead of training using only complete events, both the Topograph and \spanet models can be trained including partial events, that is, events where some partons are not able to be matched to reconstructed jets.
In the training events, for Topographs at least one $W$ boson, and for \spanet at least one top quark are required to be fully reconstructable.
For these models the efficiencies of fully reconstructable events change slightly, with Topographs having a slight reduction in efficiencies and \spanet a slight increase, as shown in \cref{tab:app_efficiencies_partial}.
However, for most selections, the efficiencies are still within the uncertainties of models trained only on fully reconstructable events.
\Cref{tab:app_correct_jets} shows the parton matching efficiencies for events categorised based on how many partons can be matched to reconstructed jets.
Here the benefit of training on partial events can be seen, with both Topographs and \spanet having higher efficiencies of correctly matching jets to the all available partons.


\begin{table*}[tb]
    \centering
    \caption{Event reconstruction efficiencies (\%) for the $\chi^2$ method, the \spanet model and our Topograph model in different jet and $b$-jet multiplicities.
    Both models were trained on complete and partial events.
    The highest efficiency is highlighted in bold.}
    \label{tab:app_efficiencies_partial}
    \begin{tabularx}{0.95\textwidth}{Y|Y|YYY} 
        \toprule
         $N_{jets}$ & $N_{b-\mathrm{jets}}$ & \spanet & Topograph & $\chi^2$ \\ 
         \midrule
         \hphantom{$\geq$ }$6$ & \hphantom{$\geq$ }$2$ & $\mathbf{81.03\pm0.32}$ & $80.77\pm0.32$ & $72.73\pm0.36$ \\
         \hphantom{$\geq$ }$6$ & $\geq2$ & $\mathbf{79.11\pm0.30}$ & $78.93\pm0.30$ & $70.94\pm0.33$ \\
         \hphantom{$\geq$ }$7$ & \hphantom{$\geq$ }$2$ & $\mathbf{65.35\pm0.44}$ & $65.18\pm0.44$ & $54.28\pm0.46$ \\
         \hphantom{$\geq$ }$7$ & $\geq2$ & $\mathbf{63.40\pm0.40}$ & $63.13\pm0.40$ & $52.11\pm0.41$ \\
         $\geq6$ & \hphantom{$\geq$ }$2$ & $\mathbf{68.93\pm0.25}$ & $68.75\pm0.25$ & $58.57\pm0.26$ \\
         $\geq6$ & $\geq2$ & $\mathbf{66.33\pm0.23}$ & $66.21\pm0.23$ & $55.90\pm0.24$ \\
         \bottomrule
    \end{tabularx}
\end{table*}

\begin{table*}[tb]
    \centering
    \caption{Percentage of events with at most $N$ incorrectly matched partons using the $\chi^2$ method, \spanet and Topographs.
    The \spanet model was trained on complete and partial events.
    Events are categorised based on the number of partons matched to jets at truth level.
    The highest efficiency is highlighted in bold.}
    \label{tab:app_correct_jets}
    \begin{tabularx}{0.95\textwidth}{YY|YYYYYY}
        \toprule
        $N_{partons}$ &  & \multicolumn{6}{c}{Incorrectly matched partons [\%]} \\
        matched & Model & 0 & $\leq1$ & $\leq2$ & $\leq3$ & $\leq4$ & $\leq5$  \\
         \midrule
         \multirow{3}{*}{3} & \spanet & $36.68\pm0.66$ & $76.94\pm0.57$ & $98.55\pm0.16$ & $100.00\pm0.00$ & - & - \\
         & Topograph & $\mathbf{37.27\pm0.66}$ & $77.89\pm0.57$ & $98.77\pm0.15$ & $100.00\pm0.00$ & - & - \\
         & $\chi^2$ & $34.30\pm0.65$ & $\mathbf{80.12\pm0.54}$ & $\mathbf{99.05\pm0.13}$ & $100.00\pm0.00$ & - & - \\
         \midrule
         \multirow{3}{*}{4} & \spanet & $37.78\pm0.28$ & $66.29\pm0.28$ & $91.78\pm0.16$ & $99.46\pm0.04$ & $100.00\pm0.00$ & -  \\
         & Topograph & $\mathbf{38.08\pm0.28}$ & $\mathbf{67.53\pm0.27}$ & $\mathbf{92.57\pm0.15}$ & $99.62\pm0.04$ & $100.00\pm0.00$ & -  \\
         & $\chi^2$ & $31.01\pm0.27$ & $63.84\pm0.28$ & $92.19\pm0.16$ & $\mathbf{99.73\pm0.03}$ & $100.00\pm0.00$ & -  \\
         \midrule
         \multirow{3}{*}{5} & \spanet & $48.14\pm0.19$ & $67.24\pm0.18$ & $88.18\pm0.12$ & $98.14\pm0.05$ & $99.95\pm0.01$ & $100.00\pm0.00$  \\
         & Topograph & $\mathbf{48.22\pm0.19}$ & $\mathbf{69.27\pm0.18}$ & $\mathbf{89.52\pm0.12}$ & $\mathbf{98.68\pm0.04}$ & $\mathbf{99.96\pm0.01}$ & $100.00\pm0.00$  \\
         & $\chi^2$ & $32.36\pm0.18$ & $58.27\pm0.19$ & $84.70\pm0.14$ & $98.21\pm0.05$ & $99.94\pm0.01$ & $100.00\pm0.00$  \\
         \midrule
         \multirow{3}{*}{6} & \spanet & $\mathbf{66.33\pm0.23}$ & $74.38\pm0.21$ & $89.86\pm0.14$ & $96.18\pm0.09$ & $99.54\pm0.03$ & $100.00\pm0.00$ \\
         & Topograph & $66.20\pm0.23$ & $\mathbf{74.87\pm0.21}$ & $\mathbf{91.11\pm0.14}$ & $\mathbf{97.04\pm0.08}$ & $\mathbf{99.73\pm0.02}$ & $100.00\pm0.00$ \\
         & $\chi^2$ & $55.90\pm0.24$ & $64.93\pm0.23$ & $84.14\pm0.17$ & $93.97\pm0.11$ & $99.43\pm0.04$ & \hphantom{$1$}$99.98\pm0.01$ \\
        \bottomrule
    \end{tabularx}
\end{table*}

\subsection{Impact of systematic variations}

For applications in high energy physics, it is crucial that any new approach is not sensitive to changes under systematic variations.
In particular with machine learning approaches, it would be problematic if methods were sensitive to underlying and non-physical effects arising from the simulated samples on which they were trained.
Other sources of variation come from differences in the calibration or reconstruction of physics objects between simulation and data.

To test the dependence on the simulated samples used to train the Topograph, we evaluate the best performing model trained on the nominal MadGraph data on an alternative independent dataset.
This alternative sample consists of all-hadronic \ttbar events simulates both the hard interactions and parton shower are with Pythia8~(v8.307), using the Monash tuned set of parameters~\cite{Monash} at leading order accuracy.

To test the dependence on reconstruction effects, we apply a shift or gaussian smearing to the energy of reconstructed jets in the events.

The absolute change in performance arising from the systematic variations is summarised in \cref{tab:systematic_uncertainties}.
The impact is compared for Topographs, \spanet, and $\chi^2$.
Evaluating on the alternative sample results in a slightly reduced overall efficiency for all three approaches.
This effects Topographs and \spanet slightly more than $\chi^2$,
however, the overall gain in performance remains similar.
Both Topographs and \spanet are robust under systematic shifts or reduced jet energy resolution, whereas the $\chi^2$ method suffers from a substantial drop in efficiency, especially at higher jet multiplicities.

\begin{table*}[tb]
    \centering
    \caption{Difference in reconstruction efficiencies when evaluating the different methods on systematic variations.
    The difference is taken with respect to \cref{tab:efficiencies}.
    The systematic variations include: evaluating on a dataset produced with a Pythia for the matrix element generation and the parton shower, evaluating on the nominal data set scaling all jet energies by 2.5\%, and evaluating on the nominal data set smearing all jet energies by 5\%.}
    \label{tab:systematic_uncertainties}
    \begin{tabularx}{0.95\textwidth}{Y|Y|YYY|YYY|YYY} 
        \toprule
         & & \multicolumn{3}{c}{Pythia} & \multicolumn{3}{c}{Jet Scale} & \multicolumn{3}{c}{Jet Resolution} \\
         $N_{jets}$ & $N_{b-\mathrm{jets}}$ & \spanet & Topograph & $\chi^2$ & \spanet & Topograph & $\chi^2$ & \spanet & Topograph & $\chi^2$ \\ 
         \midrule
         \hphantom{$\geq$ }$6$ & \hphantom{$\geq$ }$2$ & $-2.13$ & $-2.00$ & $-0.94$ & $+0.04$ & $\pm0.00$ & $\vphantom{1}-2.82$ & $+0.08$ & $\pm0.00$ & $-12.79$ \\
         \hphantom{$\geq$ }$6$ & $\geq2$ & $-2.28$ & $-2.13$ & $-1.33$ & $+0.07$ & $+0.03$ & $\vphantom{1}-3.63$ & $+0.04$ & $+0.01$ & $-14.30$ \\
         \hphantom{$\geq$ }$7$ & \hphantom{$\geq$ }$2$ & $-2.04$ & $-2.14$ & $-1.66$ & $\pm0.00$ & $\pm0.00$ & $-12.30$ & $-0.05$ & $-0.06$ & $-21.52$ \\
         \hphantom{$\geq$ }$7$ & $\geq2$ & $-2.36$ & $-2.37$ & $-1.62$ & $-0.02$ & $+0.04$ & $-12.69$ & $-0.10$ & $-0.07$ & $-21.63$ \\
         $\geq6$ & \hphantom{$\geq$ }$2$ & $-1.64$ & $-1.53$ & $-0.72$ & $+0.04$ & $+0.01$ & $\vphantom{1}-8.23$ & $+0.02$ & $+0.02$ & $-17.18$ \\
         $\geq6$ & $\geq2$ & $-1.86$ & $-1.89$ & $-0.94$ & $+0.06$ & $+0.05$ & $\vphantom{1}-9.02$ & $-0.01$ & $+0.03$ & $-17.96$ \\
         \bottomrule
    \end{tabularx}
\end{table*}

\subsection{Studying edge scores}

\label{app:studies}

\begin{figure*}[htbp]
    \centering
    \begin{subfigure}[b]{0.45\textwidth}
        \includegraphics[width=\textwidth]{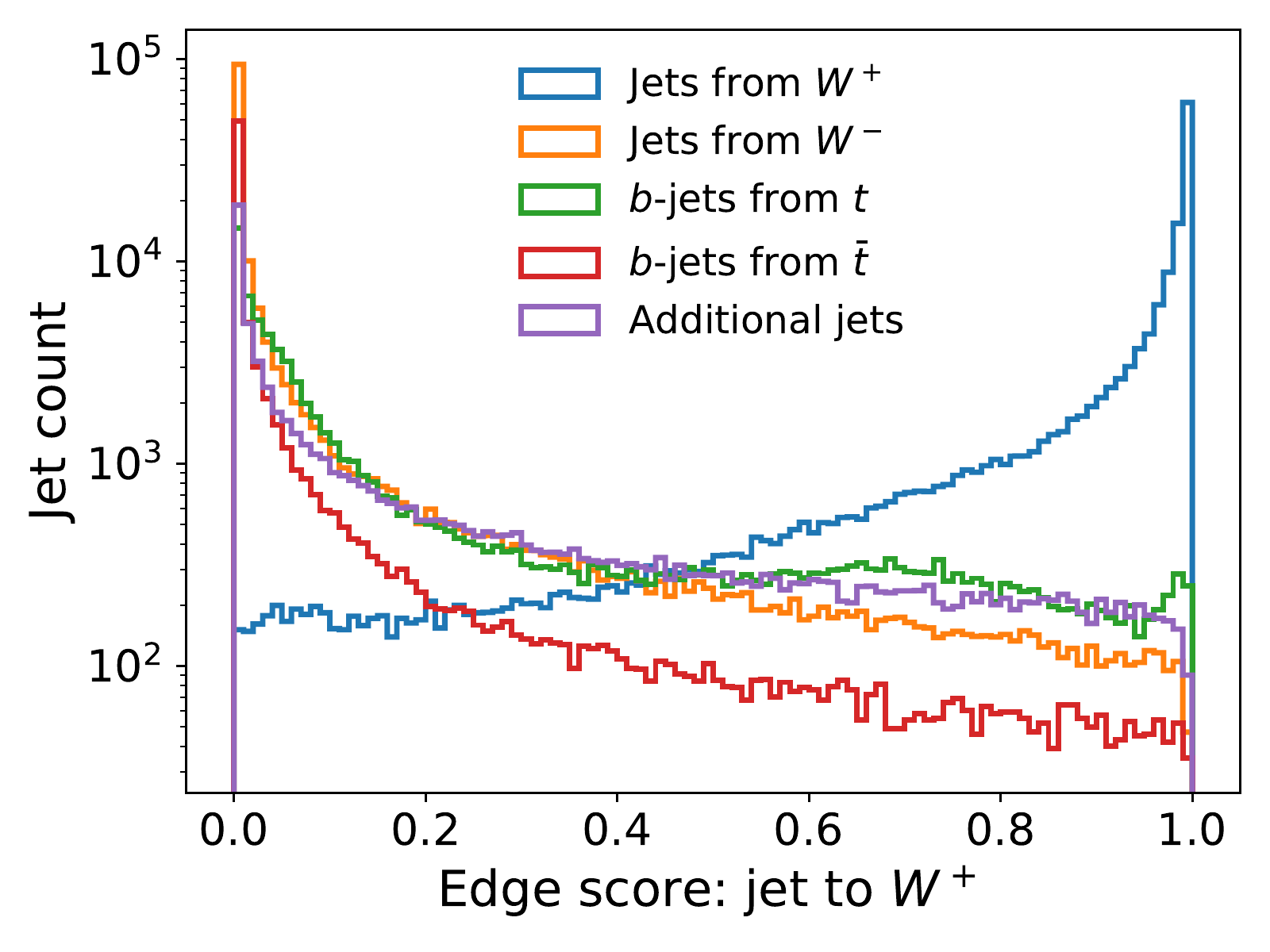}
        \caption{ }
        \label{fig:app_scores_wplus:all}
    \end{subfigure}    
    \begin{subfigure}[b]{0.45\textwidth}
        \includegraphics[width=\textwidth]{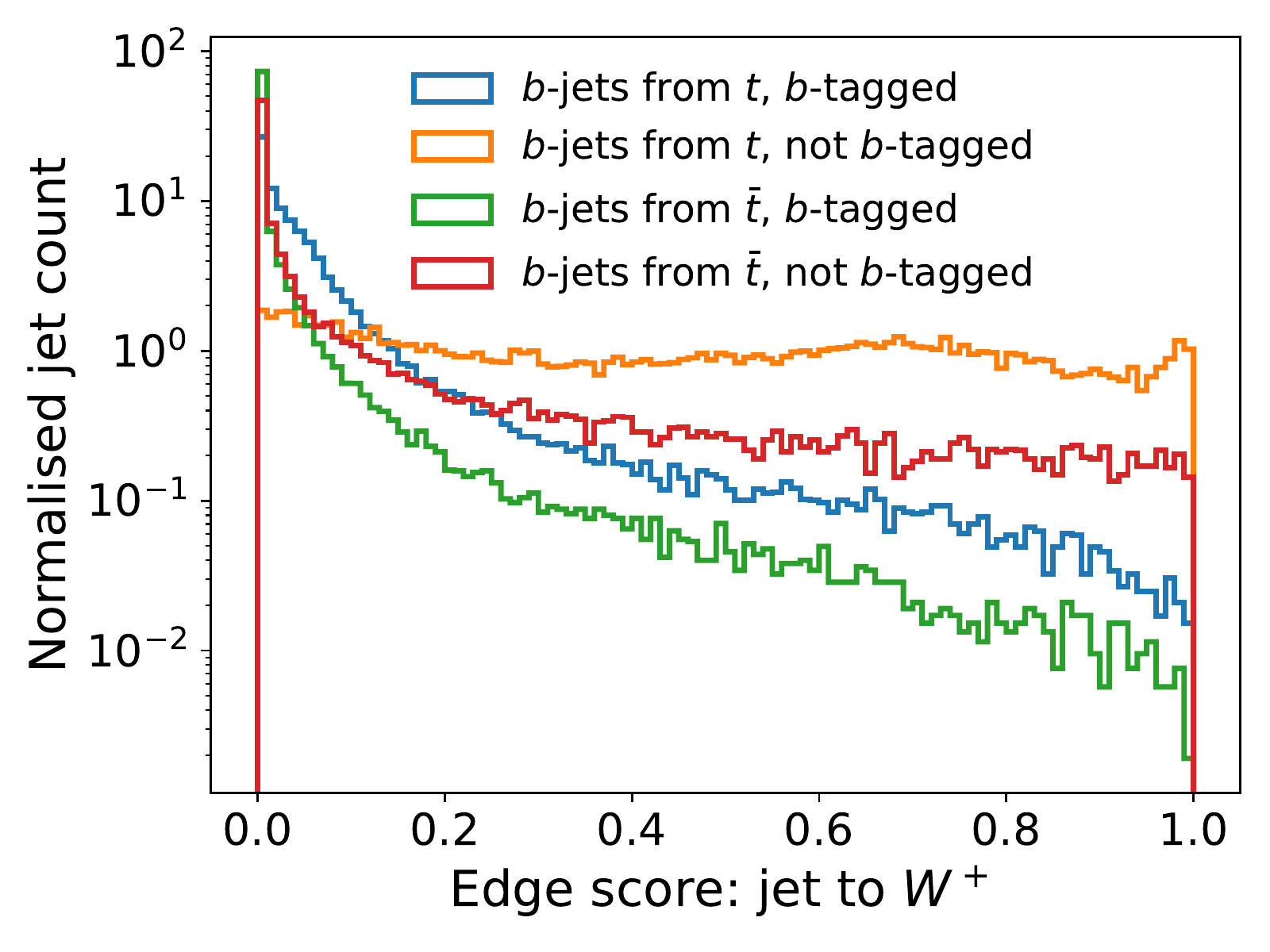}
        \caption{ }
        \label{fig:app_scores_wplus:btag}
    \end{subfigure}    
    \caption{Edge scores of jets to the helper node representing the $W^+$.
    The decision which $W$ node is the $W^+$ is taken by choosing the minimum of the loss under both hypotheses.
    \subref{fig:app_scores_wplus:all} shows all jets, while \subref{fig:app_scores_wplus:btag} only shows the $b$-jets from the two tops.
    They are further split into whether the jet was tagged as a $b$-jet or not.
    No requirement is placed on the number of $b$-tags.}
    \label{fig:app_scores_wplus}
\end{figure*}

\cref{fig:app_scores_wplus} shows the distribution of edge scores of the jets to the $W$ helper node which is decided to be the $W^+$ based on the loss.
\cref{fig:app_scores_wplus:all} includes all jets in the event.
The distribution for the jets which originate from the $W^+$ peak at one, whereas the distributions of all other jet types peak at zero.
A small peak at one can be seen for the $b$-jets originating from the top quark.
To investigate this peak, the scores of the $b$-jets from both the top and the anti-top quark are shown in \cref{fig:app_scores_wplus:btag}.
They are further split based on their $b$-tagging score.
The small peak at one originates from $b$-jets from the top quark which are not $b$-tagged.
Furthermore, the impact of the $b$-tagging on the score can be seen.
The $b$-jets from the anti-top quark which are not $b$-tagged have on average higher scores than the $b$-jets from the top quark which are $b$-tagged.
This could point to the $b$-tagging result being more important for the association to the $W$ nodes than kinematics.

\begin{figure*}[htbp]
    \centering
    \begin{subfigure}[b]{0.45\textwidth}
        \includegraphics[width=\textwidth]{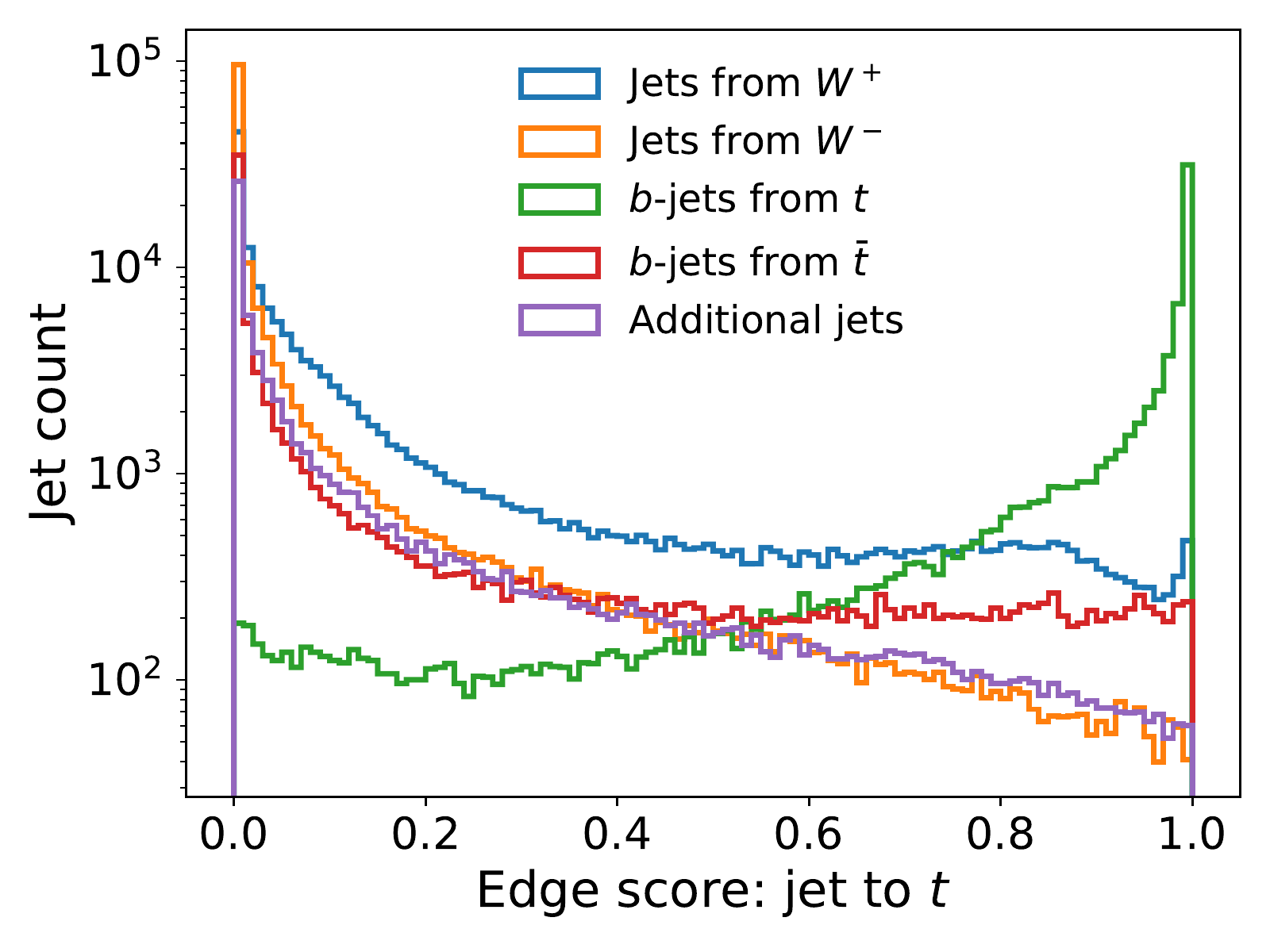}
        \caption{ }
        \label{fig:app_scores_top:all}
    \end{subfigure}        
    \begin{subfigure}[b]{0.45\textwidth}
        \includegraphics[width=\textwidth]{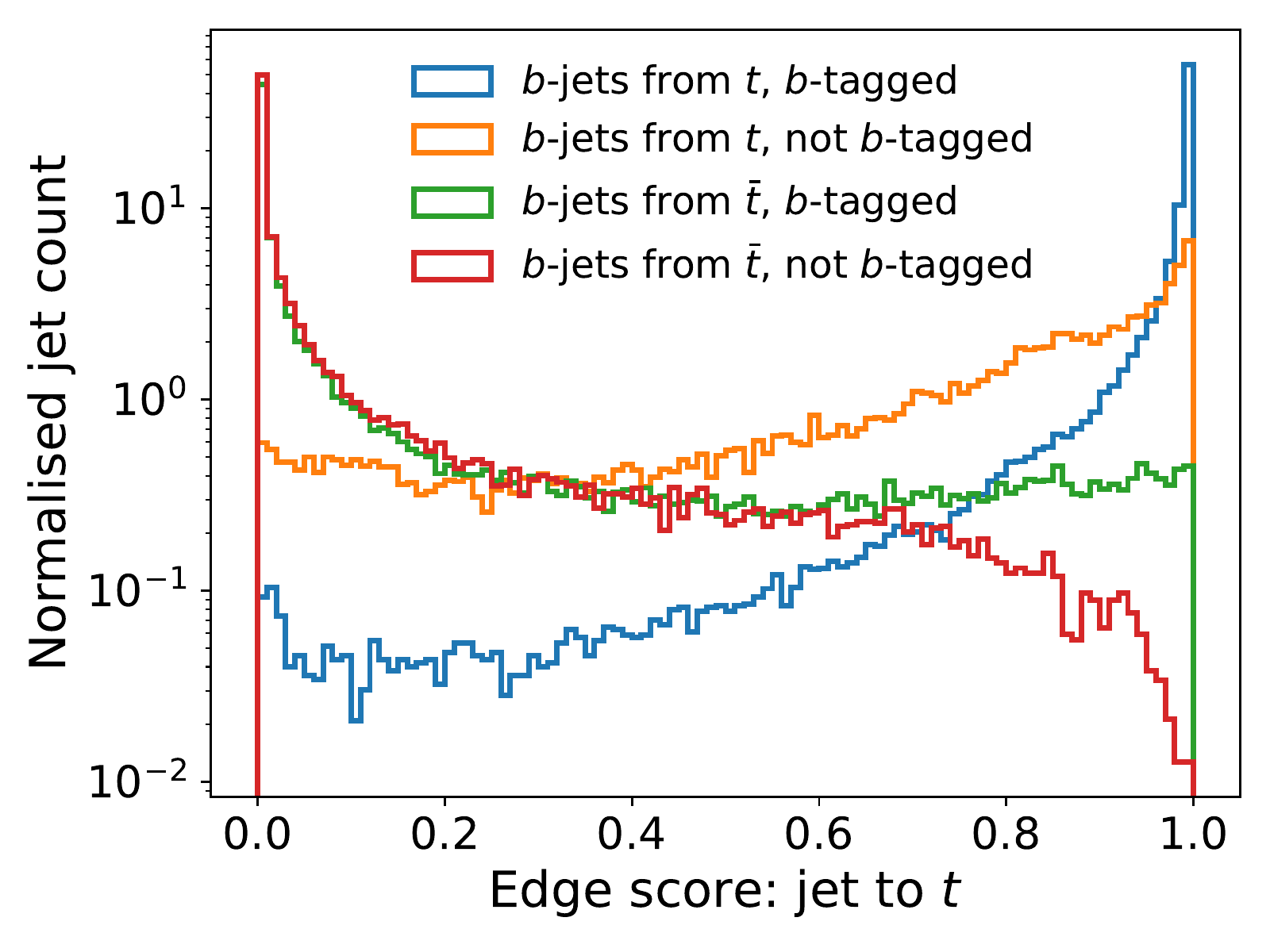}
        \caption{ }
        \label{fig:app_scores_top:btag}
    \end{subfigure}       
    \caption{Edge scores of jets to the helper node representing the $t$.
    The decision which $t$ node is the $t$ is taken by choosing the minimum of the loss under both hypotheses.
    \subref{fig:app_scores_top:all} shows all jets, while \subref{fig:app_scores_top:btag} only shows the $b$-jets from the two tops.
    They are further split into whether the jet was tagged as a $b$-jet or not.
    No requirement is placed on the number of $b$-tags.}
    \label{fig:app_scores_top}
\end{figure*}

\cref{fig:app_scores_top} shows the same distributions but to the top helper node which is decided to be the top quark.
Again, good separation of the true edges from the false edges can be seen.
Considering only $b$-jets and splitting them based on their $b$-tagging score, it can be seen, that the $b$-tagged jets originating from the anti-top quark have on average lower scores than the non $b$-tagged jets originating from the top quark.
So, the $b$-tagging decision is not as important for the association to the top helper nodes.

This reliance on the $b$-tagging result is not unexpected.
With a $b$-tagging efficiency of $70\%$, around $30\%$ of the $b$-jets will not be tagged as a $b$-jet, leaving a large contribution of true edges to the top helper nodes which are not $b$-tagged.
The observed mis-tag efficiency is around $5\%$. 
Therefore, only a small fraction of the true edges to the $W$ helper nodes is $b$-tagged.

\begin{figure*}[htbp]
    \centering
    \begin{subfigure}[b]{0.45\textwidth}
        \includegraphics[width=\textwidth]{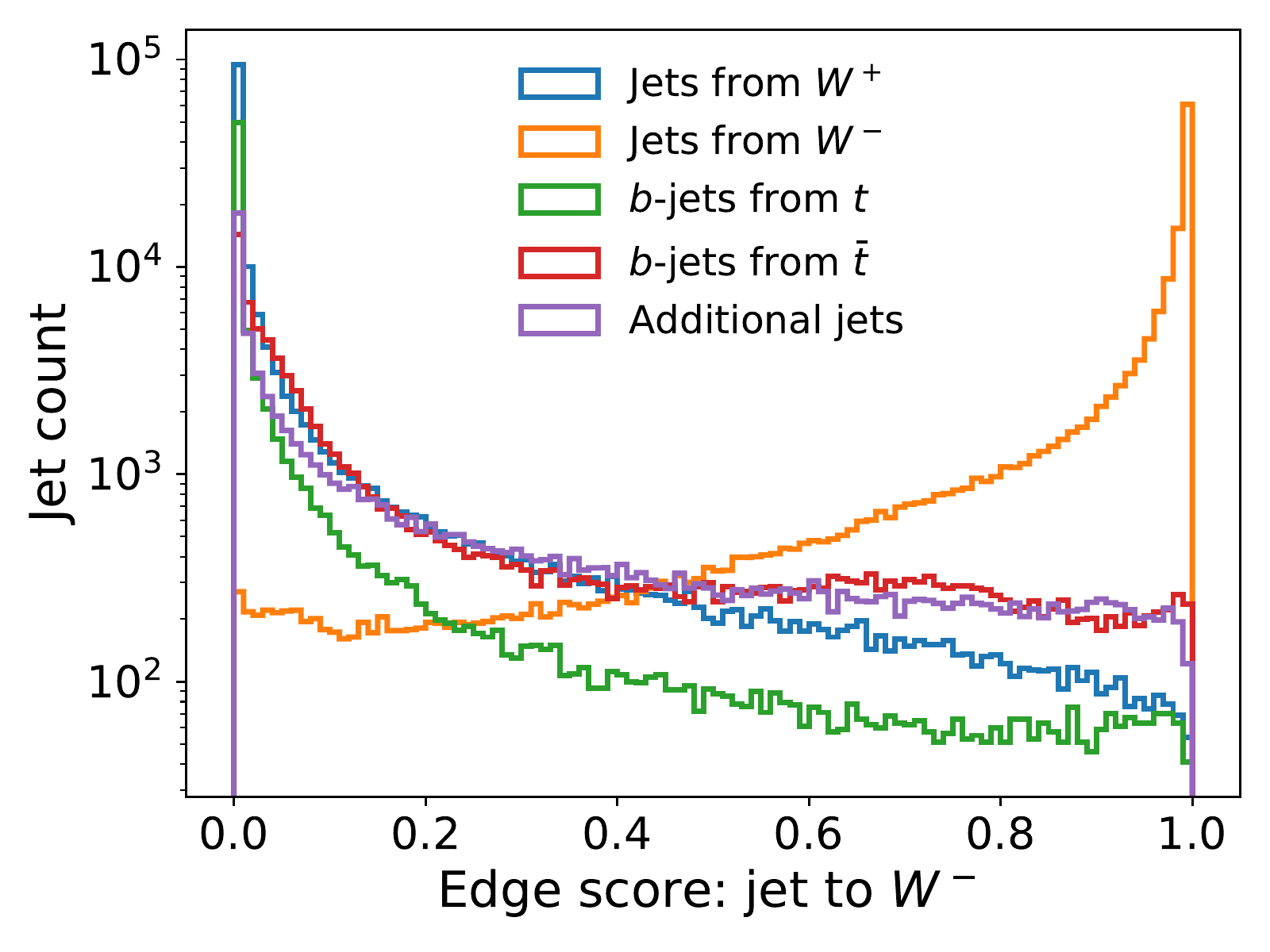}
        \caption{ }
        \label{fig:app_scores_wminus:all}
    \end{subfigure}    
    \begin{subfigure}[b]{0.45\textwidth}
        \includegraphics[width=\textwidth]{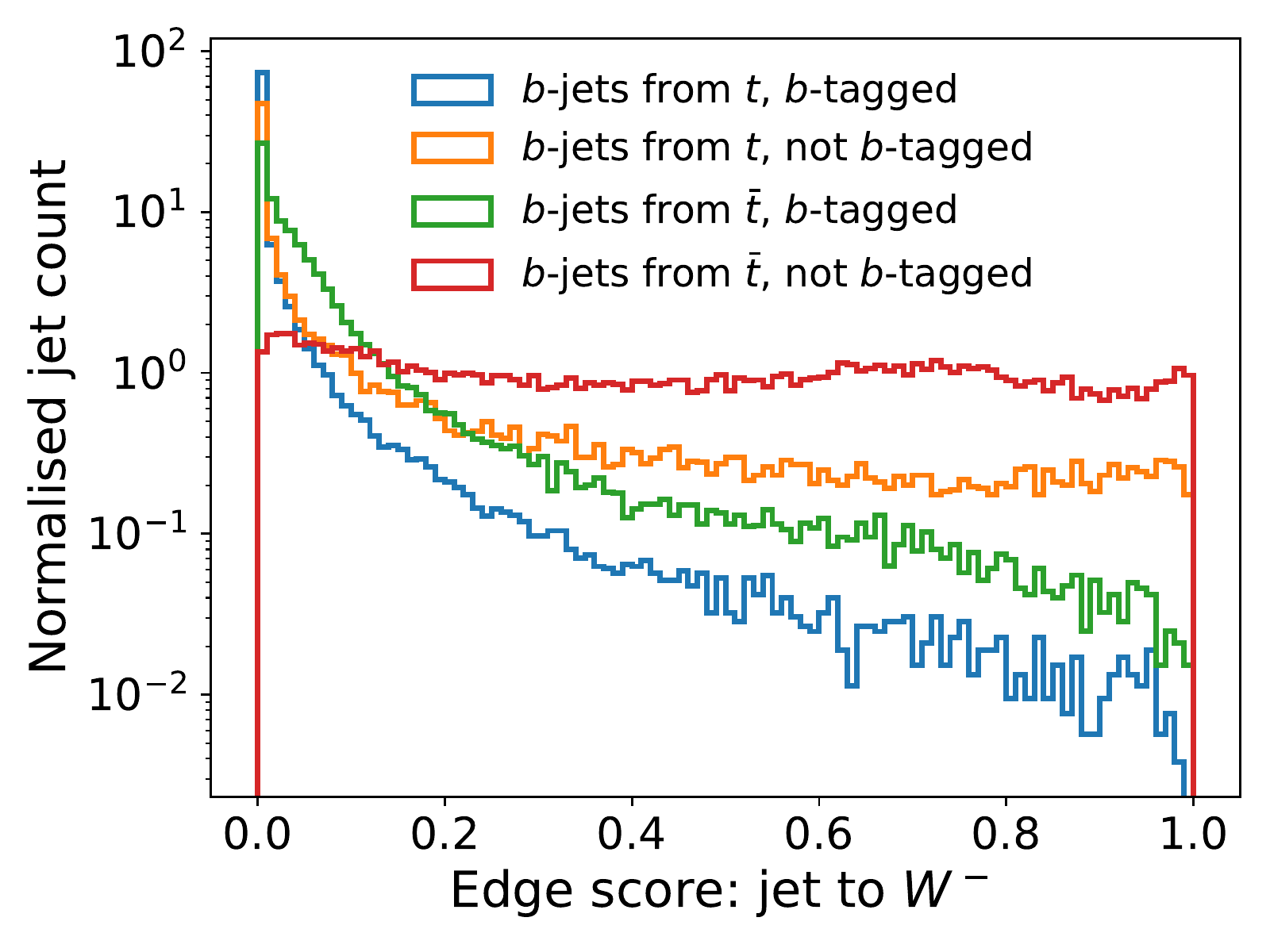}
        \caption{ }
        \label{fig:app_scores_wminus:btag}
    \end{subfigure}        
    \caption{Edge scores of jets to the helper node representing the $W^-$.
    The decision which $W$ node is the $W^-$ is taken by choosing the minimum of the loss under both hypotheses.
    \subref{fig:app_scores_wminus:all} shows all jets, while \subref{fig:app_scores_wminus:btag} only shows the $b$-jets from the two tops.
    They are further split into whether the jet was tagged as a $b$-jet or not.
    No requirement is placed on the number of $b$-tags.}
    \label{fig:app_scores_wminus}
\end{figure*}

\begin{figure*}[htbp]
    \centering
    \begin{subfigure}[b]{0.45\textwidth}
        \includegraphics[width=\textwidth]{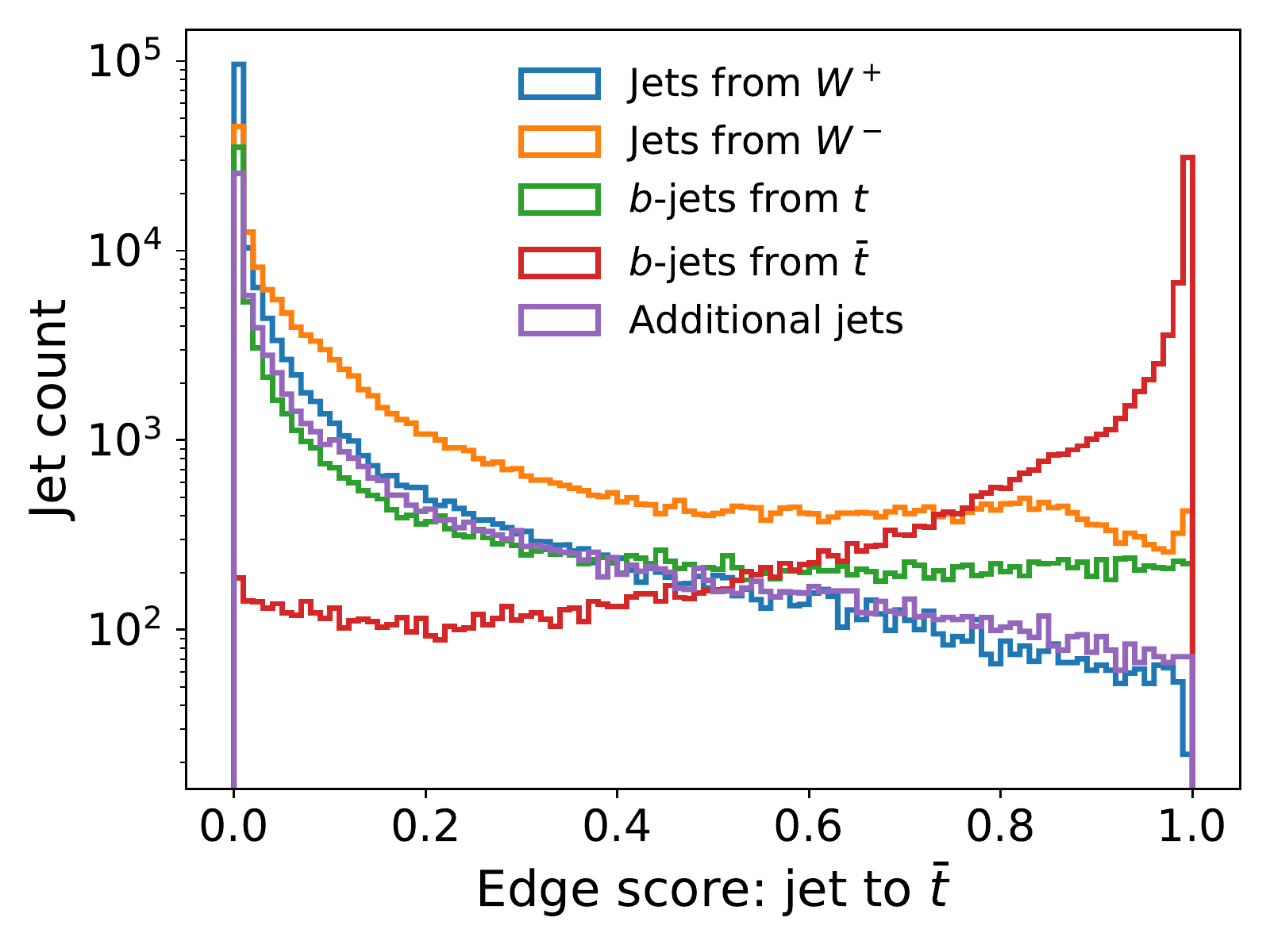}
        \caption{ }
        \label{fig:app_scores_antitop:all}
    \end{subfigure}        
    \begin{subfigure}[b]{0.45\textwidth}
        \includegraphics[width=\textwidth]{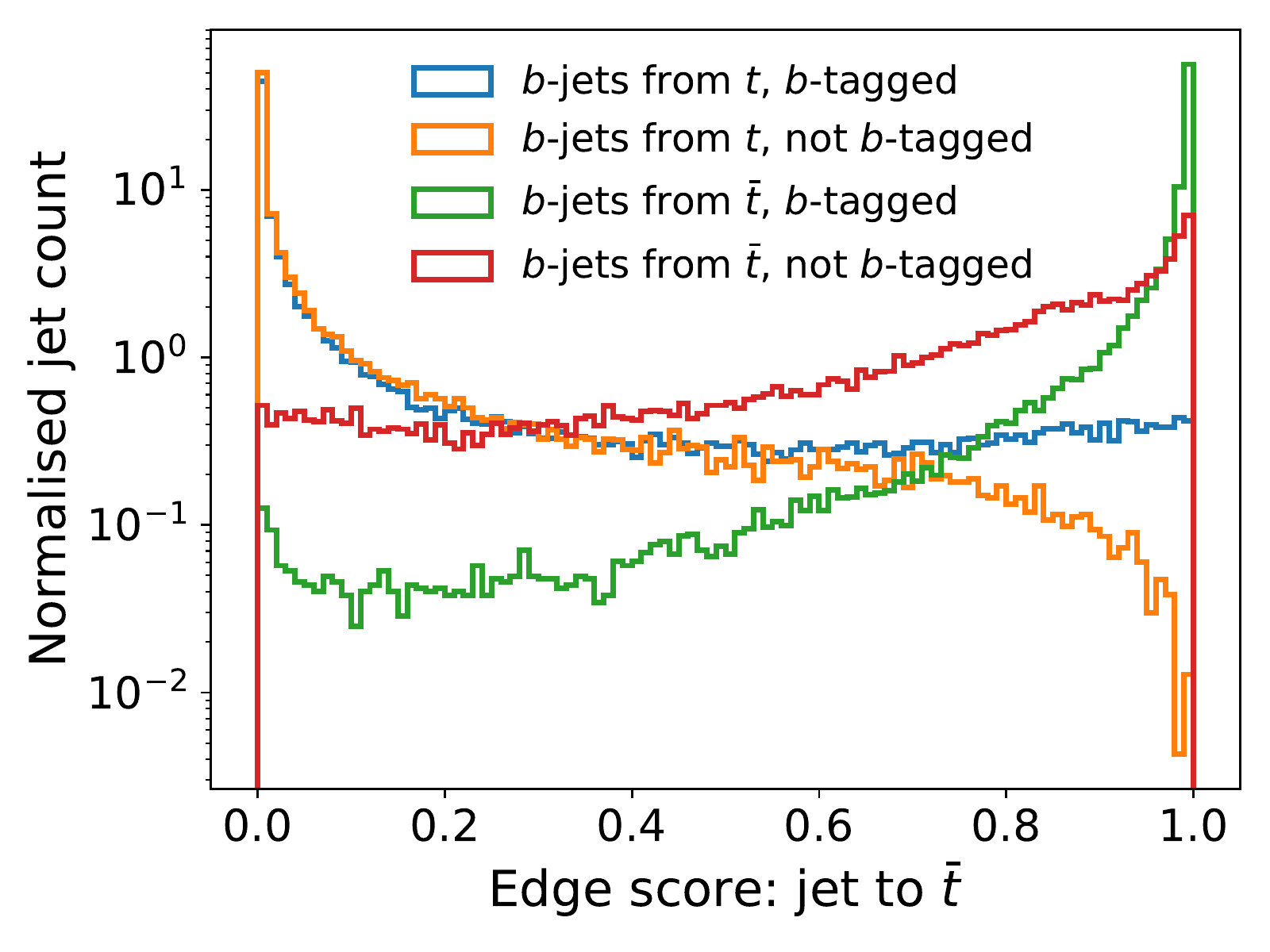}
        \caption{ }
        \label{fig:app_scores_antitop:btag}
    \end{subfigure}           
    \caption{Edge scores of jets to the helper node representing the $\bar{t}$.
    The decision which $t$ node is the $\bar{t}$ is taken by choosing the minimum of the loss under both hypotheses.
    \subref{fig:app_scores_antitop:all} shows all jets, while \subref{fig:app_scores_antitop:btag} only shows the $b$-jets from the two tops.
    They are further split into whether the jet was tagged as a $b$-jet or not.
    No requirement is placed on the number of $b$-tags.}
    \label{fig:app_scores_antitop}
\end{figure*}

\cref{fig:app_scores_wminus,fig:app_scores_antitop} show the same plots for the $W^-$ and anti-top. 
No qualitative differences can be observed to the plots for the $W^+$ and the top.

\subsection{Additional figures}
\label{app:plots}

\begin{figure*}[hbt]
    \centering
    \begin{subfigure}[b]{0.45\textwidth}
        \includegraphics[width=\textwidth]{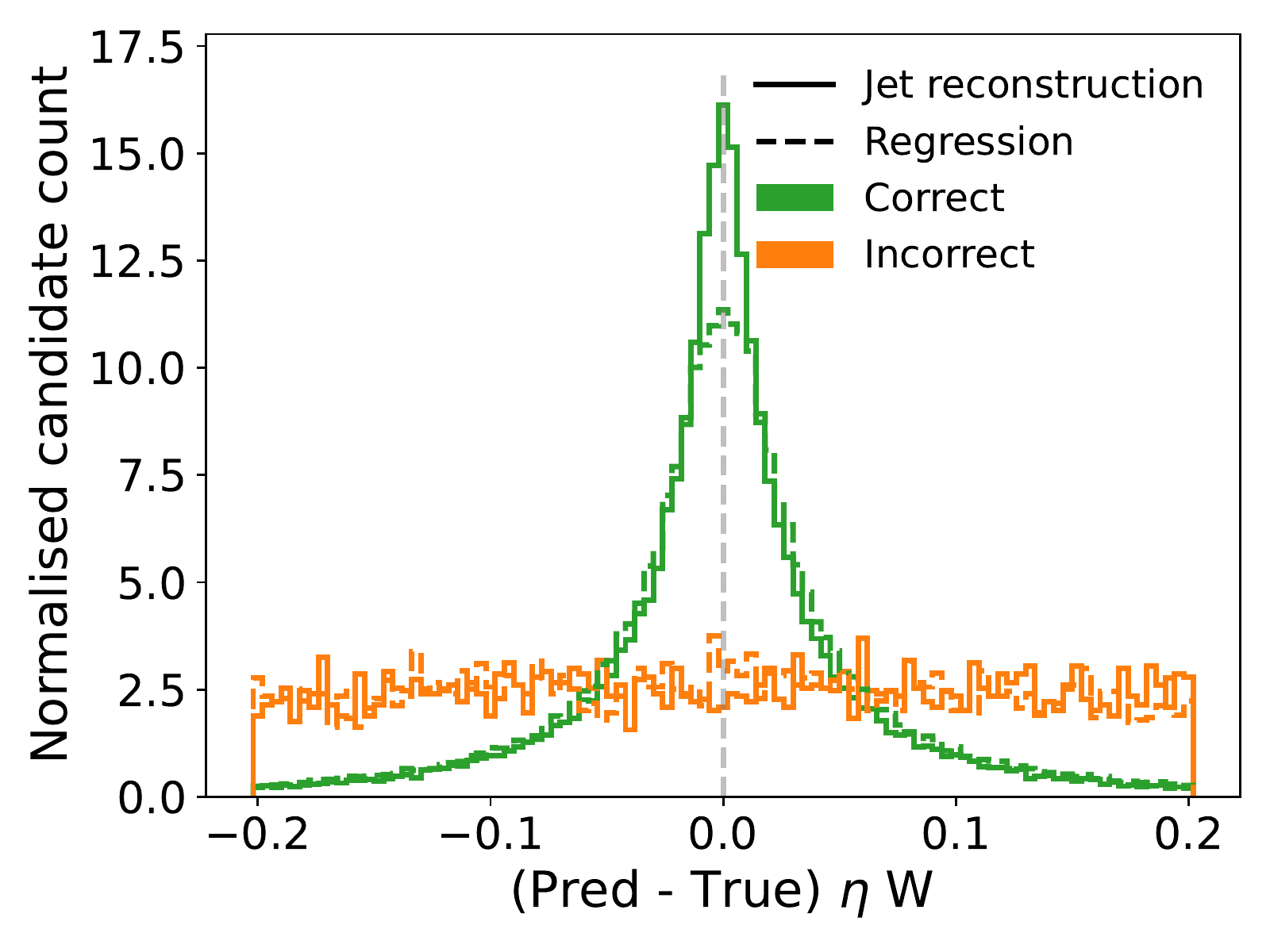}
        \caption{ }
        \label{fig:app_regression_w:eta}
    \end{subfigure}
    \begin{subfigure}[b]{0.45\textwidth}
        \includegraphics[width=\textwidth]{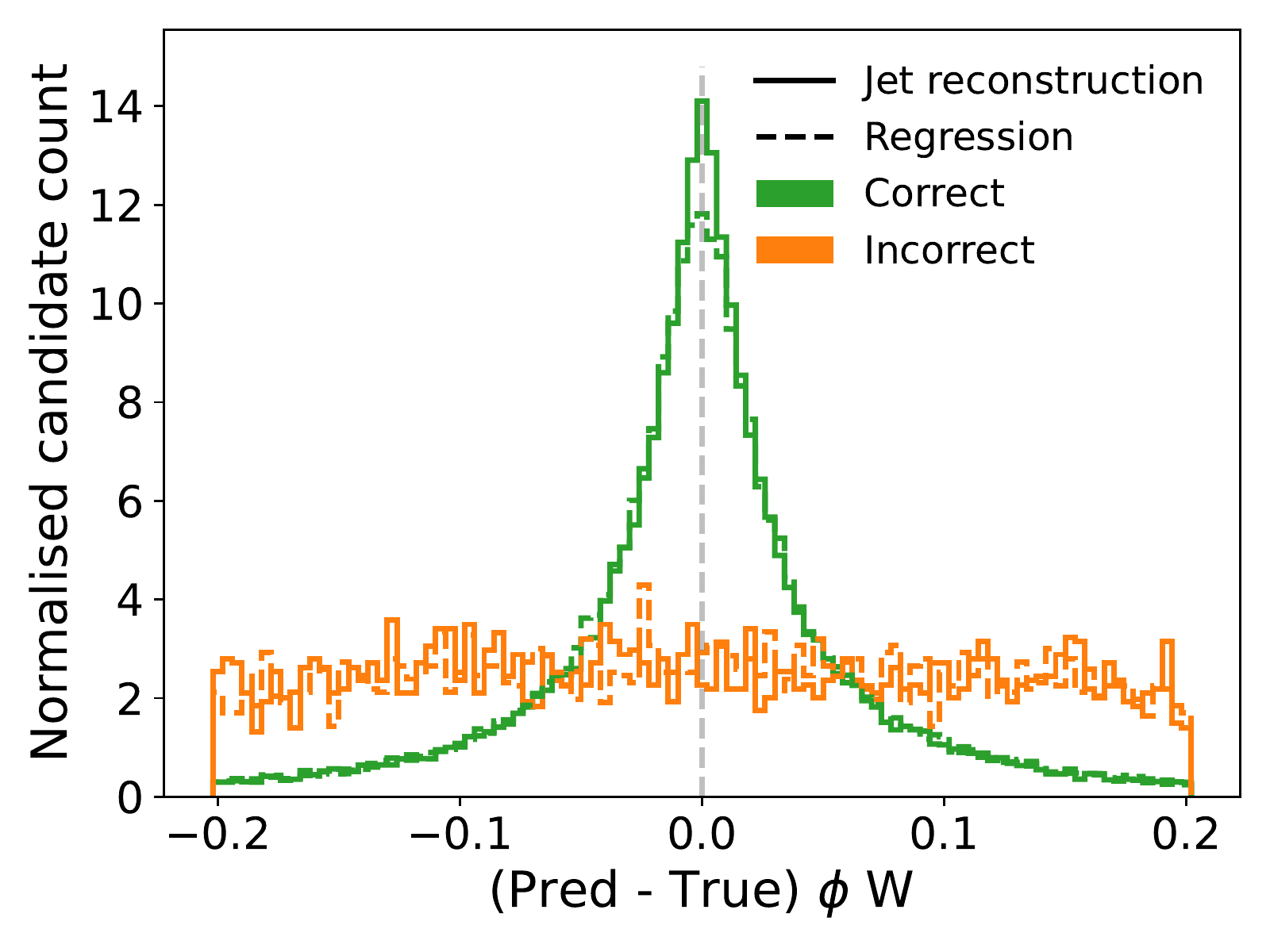}
        \caption{ }
        \label{fig:app_regression_w:phi}
    \end{subfigure}                         
    \caption{Resolution of the reconstructed $W$ boson $\eta$ and $\phi$ coordinates from the Topograph.
    Comparing the prediction from the invariant system of the assigned jets (solid line) and the Topograph regression network (dashed line) for correct assigned events (green) and incorrect assigned events (orange).}
    \label{fig:app_regression_w}
\end{figure*}

\begin{figure*}[hbt]
    \centering
    \begin{subfigure}[b]{0.45\textwidth}
        \includegraphics[width=\textwidth]{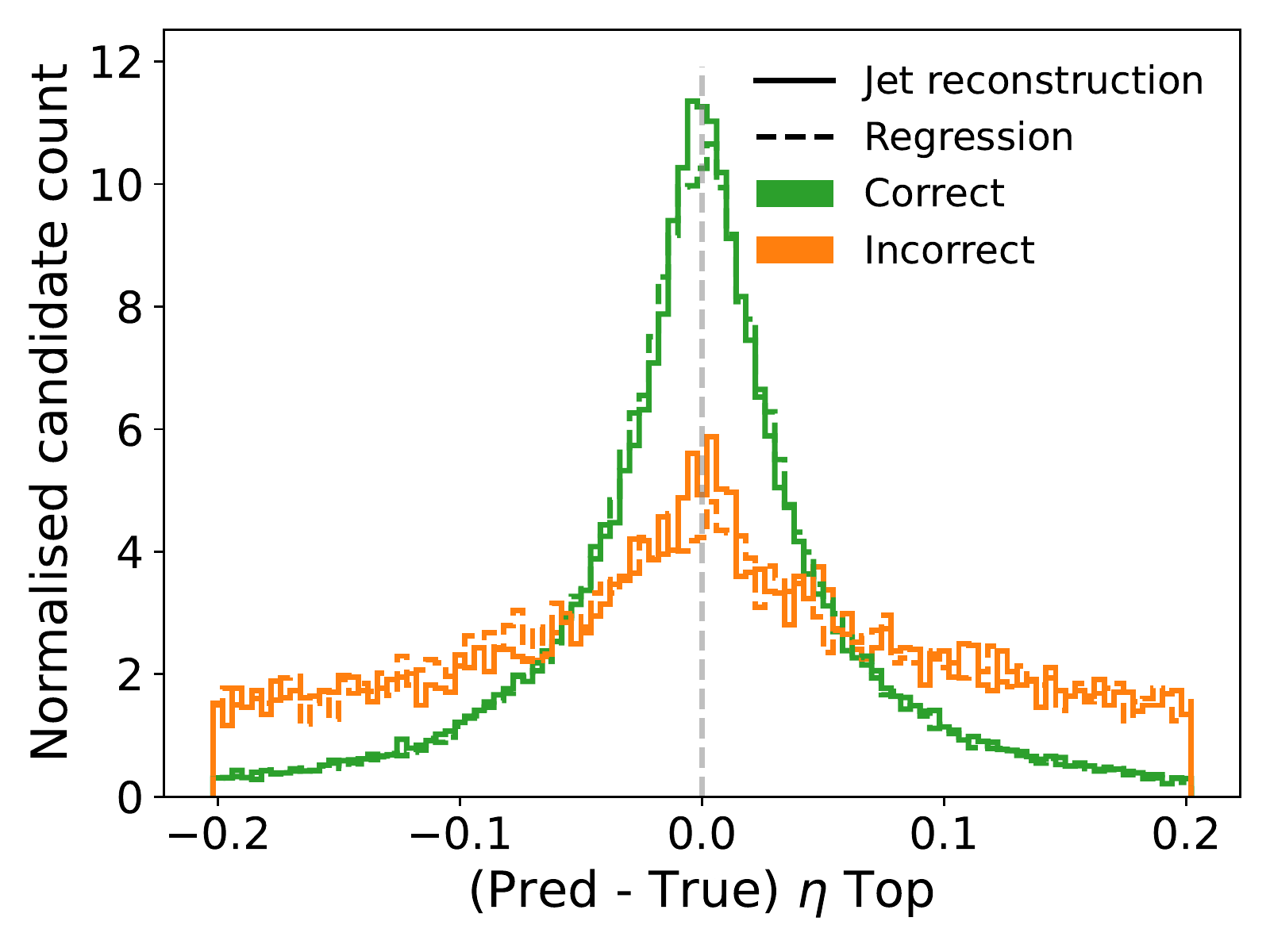}
        \caption{ }
        \label{fig:app_regression_top:eta}
    \end{subfigure}
    \begin{subfigure}[b]{0.45\textwidth}
        \includegraphics[width=\textwidth]{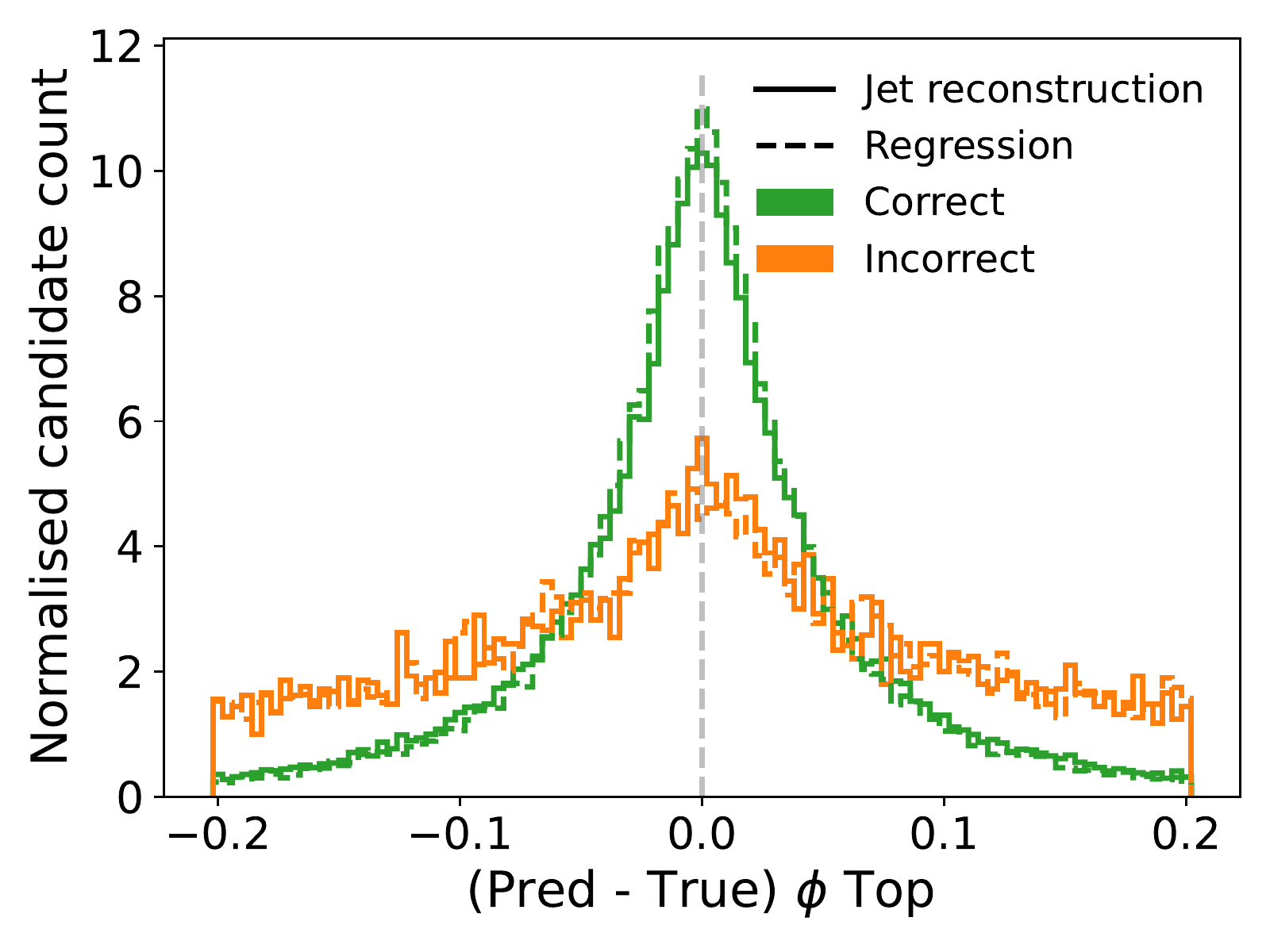}
        \caption{ }
        \label{fig:app_regression_top:phi}
    \end{subfigure}      
    \caption{Resolution of the reconstructed top quark $\eta$ and $\phi$ coordinates from the Topograph.
    Comparing the prediction from the invariant system of the assigned jets (solid line) and the Topograph regression network (dashed line) for correct assigned events (green) and incorrect assigned events (orange).}
    \label{fig:app_regression_top}
\end{figure*}

\Cref{fig:app_regression_w,fig:app_regression_top} show the regression results for the $\eta$ and $\phi$ coordinate. 
The difference between the value of the predicted $\eta$ or $\phi$ and the true parton property is shown.
For the predicted properties, the parton is either reconstructed from the jets that the model associates to the parton or the regression result is taken.
`Correct' and `incorrect' events are shown in separate distributions. 
For `incorrect' events an additional prediction is shown by taking the true jets to reconstruct the parton properties.
For `correct' events, the regression has for both quantities a wider distribution than the model association. 
For `incorrect' events, the regression and the model association perform worse, however using the true jets also has a worse resolution for these events.
Especially for the $W$ boson, the regression and the model association have a worse performance.
However, for the top quark `incorrect' labelled events can still contain the correct three jets but with a wrong assignment by switching one of the $W$-jets with the $b$-jet.

\begin{figure*}[hbt]
    \centering
    \begin{subfigure}[b]{0.32\textwidth}
        \includegraphics[width=\textwidth]{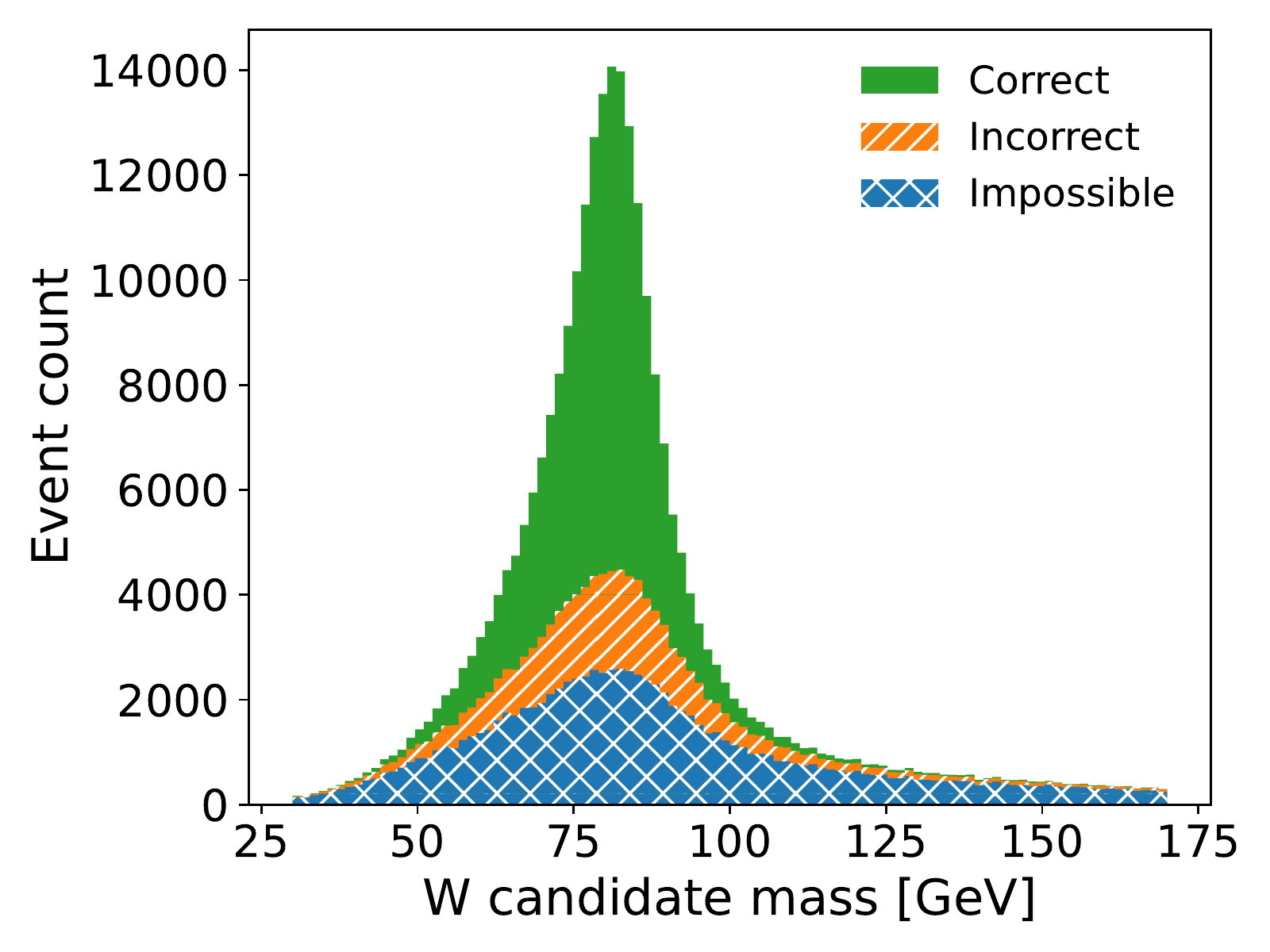}
        \caption{ }
        \label{fig:app_stacked_reco_mass_w:topo}
    \end{subfigure}
    \begin{subfigure}[b]{0.32\textwidth}
        \includegraphics[width=\textwidth]{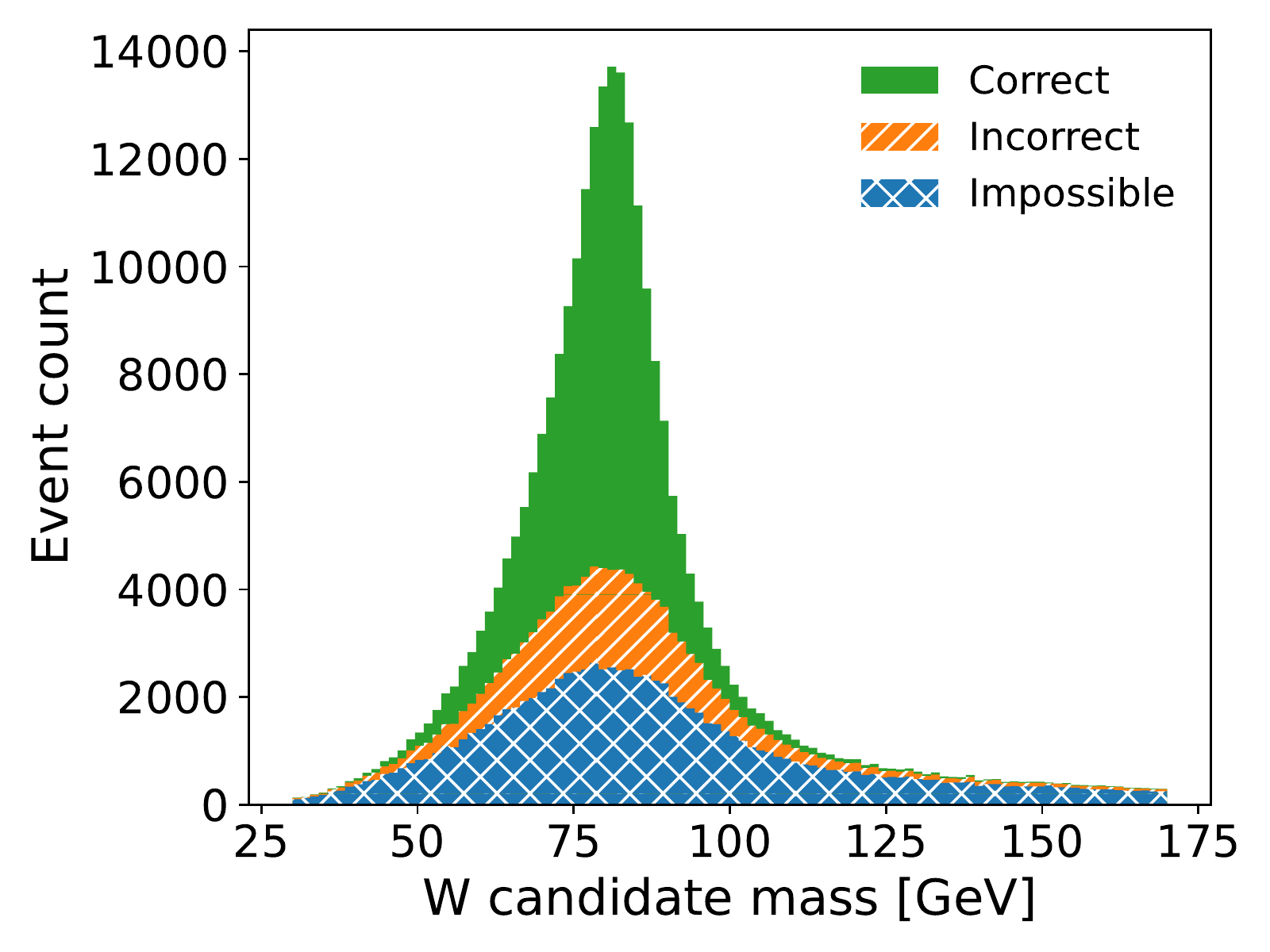}
        \caption{ }
        \label{fig:app_stacked_reco_mass_w:spa}
    \end{subfigure} 
    \begin{subfigure}[b]{0.32\textwidth}
        \includegraphics[width=\textwidth]{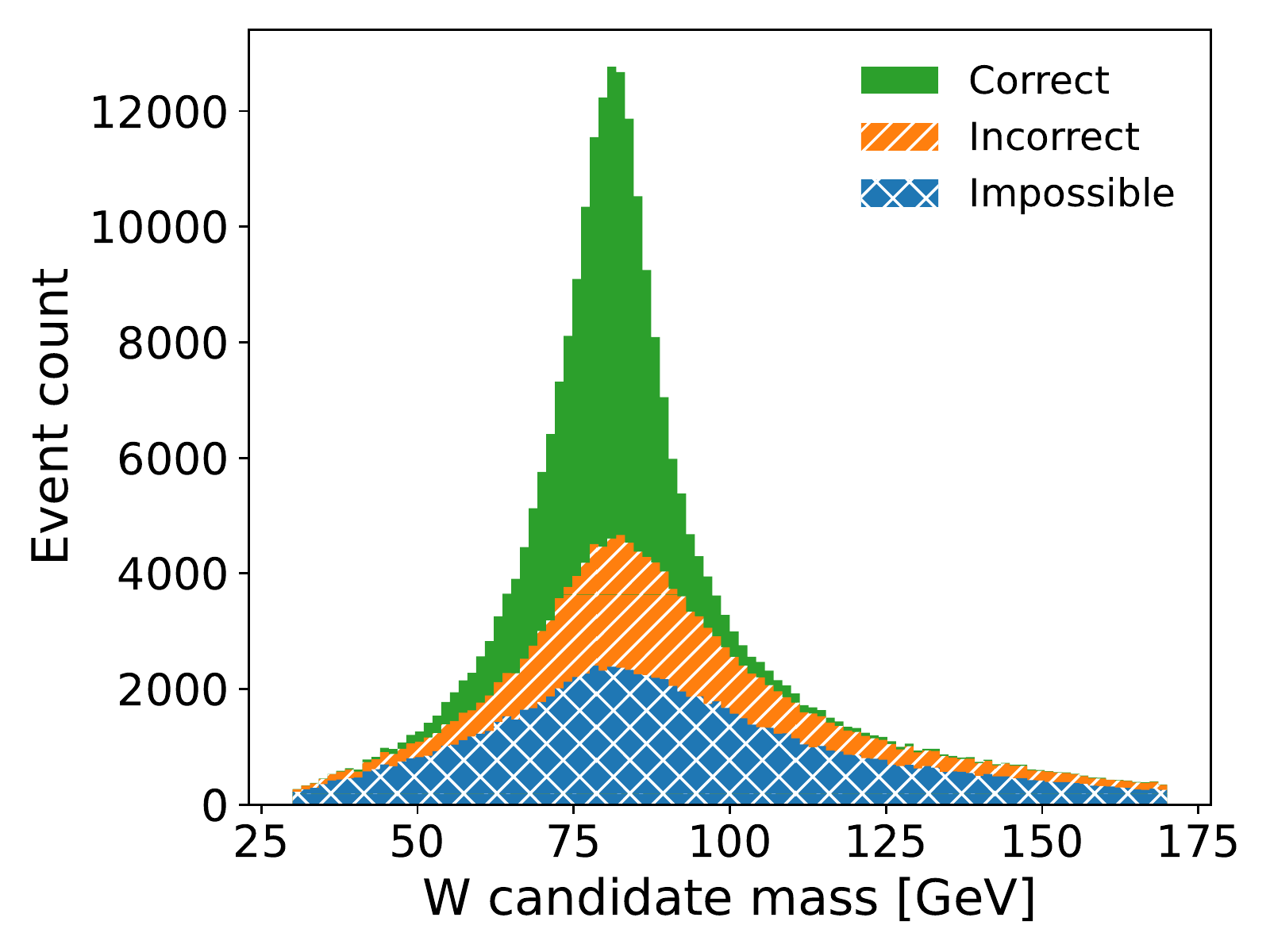}
        \caption{ }
        \label{fig:app_stacked_reco_mass_w:chi2}
    \end{subfigure}             
    \caption{Stacked distributions of the reconstructed $m_W$ using \subref{fig:app_stacked_reco_mass_w:topo} Topographs, \subref{fig:app_stacked_reco_mass_w:spa}\spanet, and \subref{fig:app_stacked_reco_mass_w:chi2} $\chi^2$.
    The single histograms split the candidates into `correct', `incorrect', and `impossible' candidates.}
    \label{fig:app_stacked_reco_mass_w}
\end{figure*}

\begin{figure*}[hbt]
    \centering
    \begin{subfigure}[b]{0.32\textwidth}
        \includegraphics[width=\textwidth]{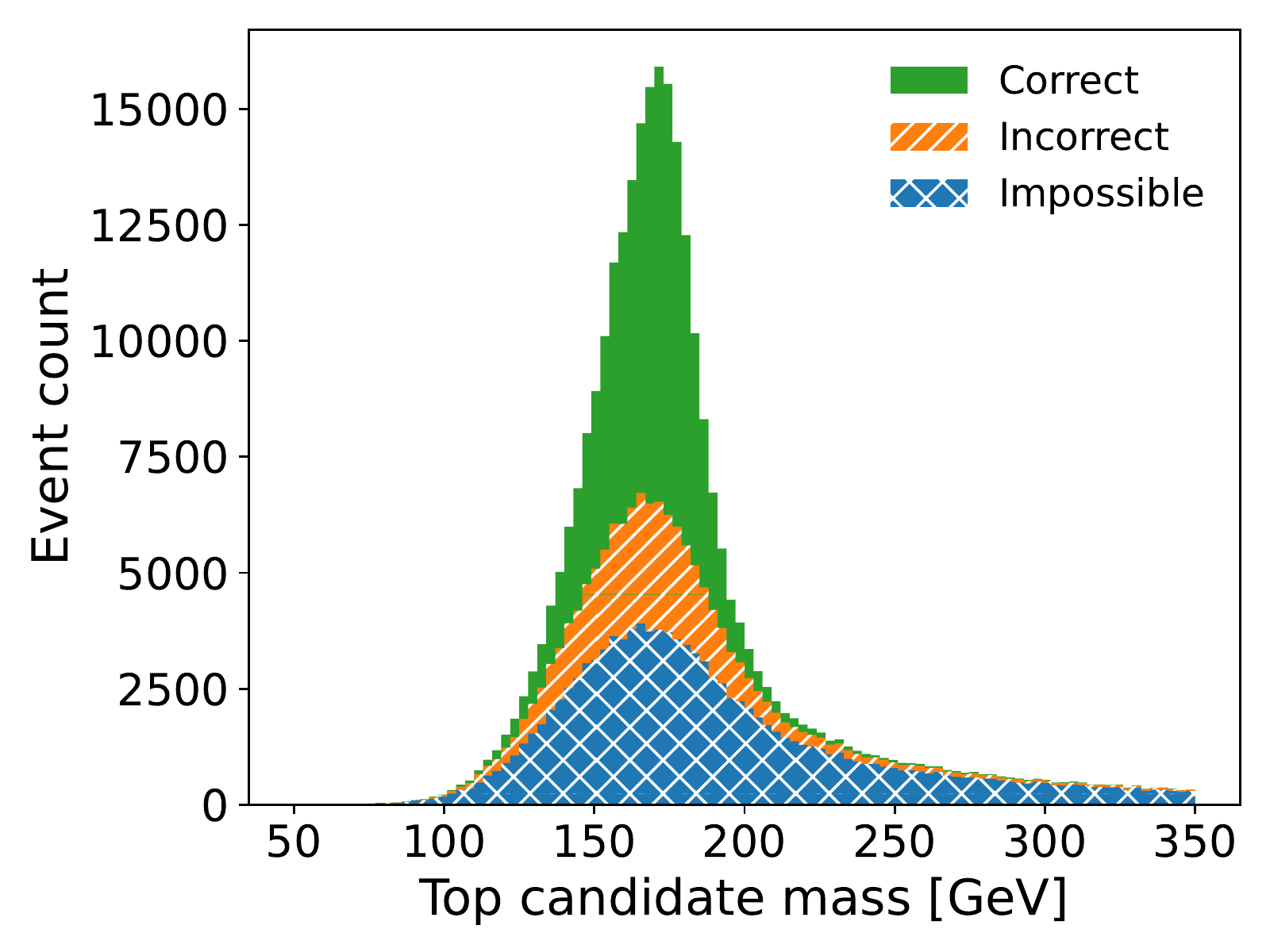}
        \caption{ }
        \label{fig:app_stacked_reco_mass_top:topo}
    \end{subfigure}
    \begin{subfigure}[b]{0.32\textwidth}
        \includegraphics[width=\textwidth]{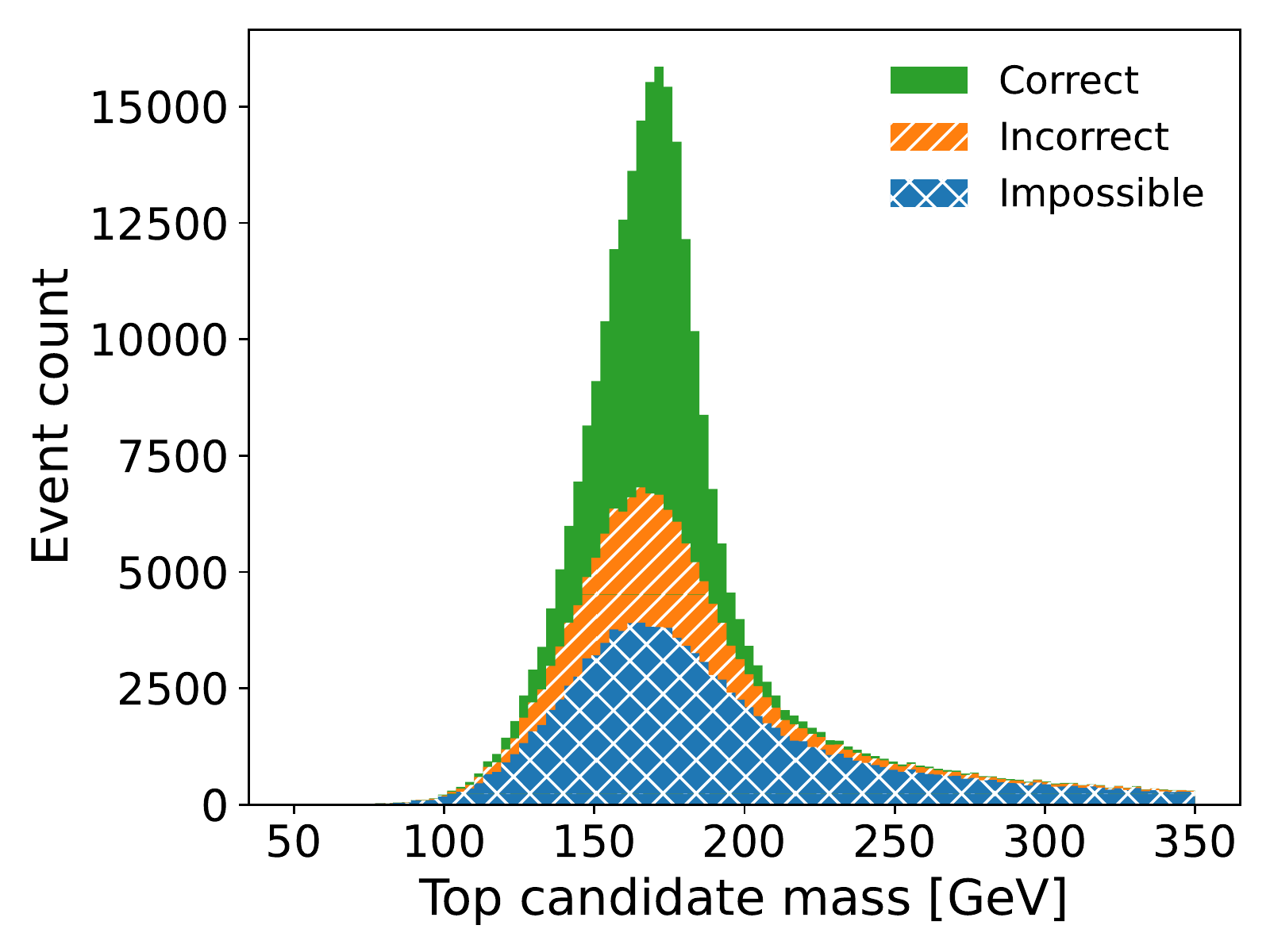}
        \caption{ }
        \label{fig:app_stacked_reco_mass_top:spa}
    \end{subfigure}
    \begin{subfigure}[b]{0.32\textwidth}
        \includegraphics[width=\textwidth]{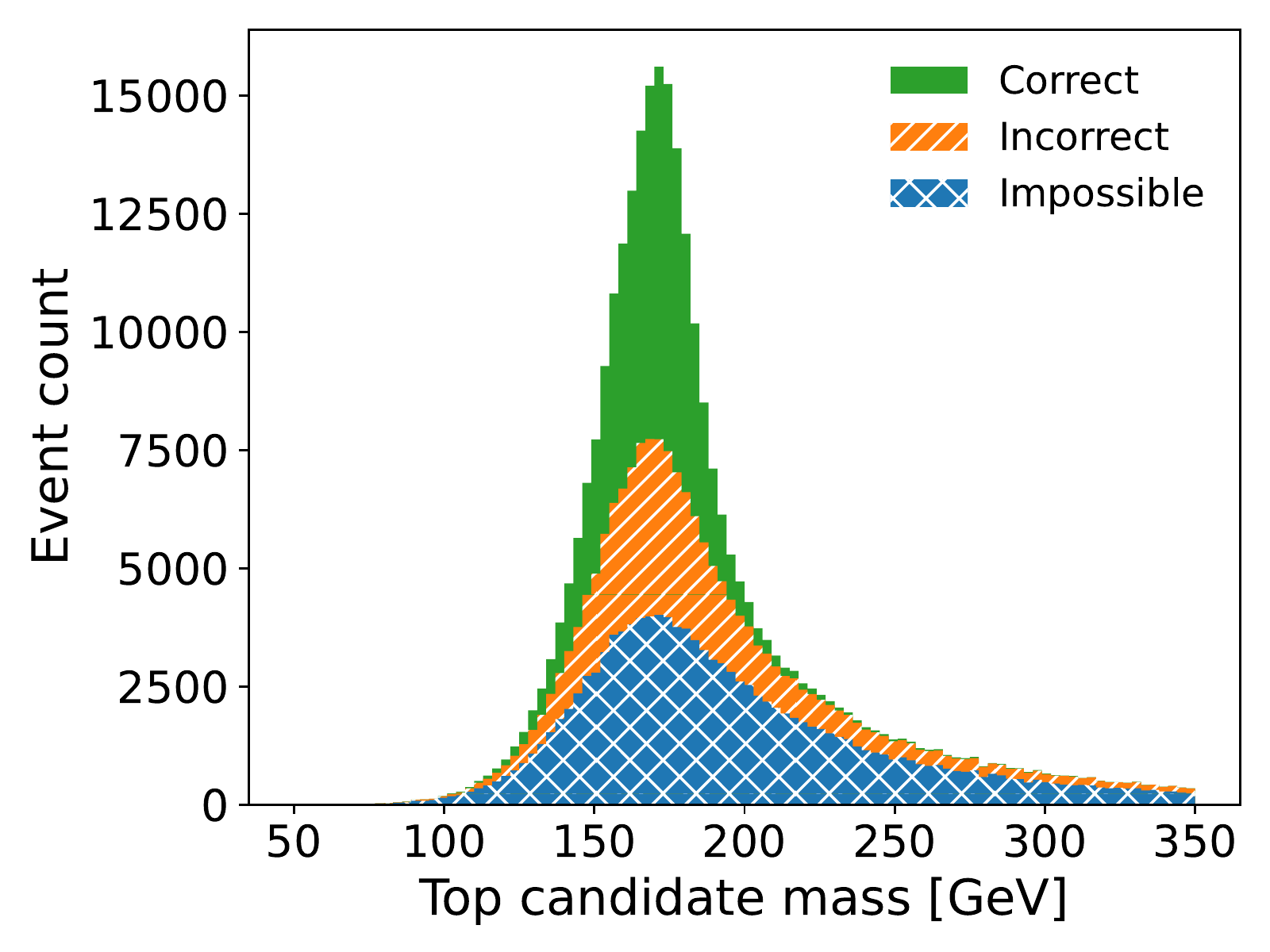}
        \caption{ }
        \label{fig:app_stacked_reco_mass_top:chi2}
    \end{subfigure}            
    \caption{Stacked distributions of the reconstructed $m_{top}$ using \subref{fig:app_stacked_reco_mass_top:topo} Topographs, \subref{fig:app_stacked_reco_mass_top:spa}\spanet, and \subref{fig:app_stacked_reco_mass_top:chi2} $\chi^2$.
    The single histograms split the candidates into `correct', `incorrect', and `impossible' candidates.}
    \label{fig:app_stacked_reco_mass_top}
\end{figure*}

\Cref{fig:app_stacked_reco_mass_w} shows the distribution of the reconstructed $W$ mass split into `correct', `incorrect', and `impossible' events as stacked histograms as an alternative representation compared to \cref{fig:reco_mass_comparison_w}. Similarly, \cref{fig:app_stacked_reco_mass_top} is an alternative representation of \cref{fig:reco_mass_comparison_top} showing the reconstructed top quark mass.

\begin{figure*}[hbt]
    \centering
    \begin{subfigure}[b]{0.32\textwidth}
        \includegraphics[width=\textwidth]{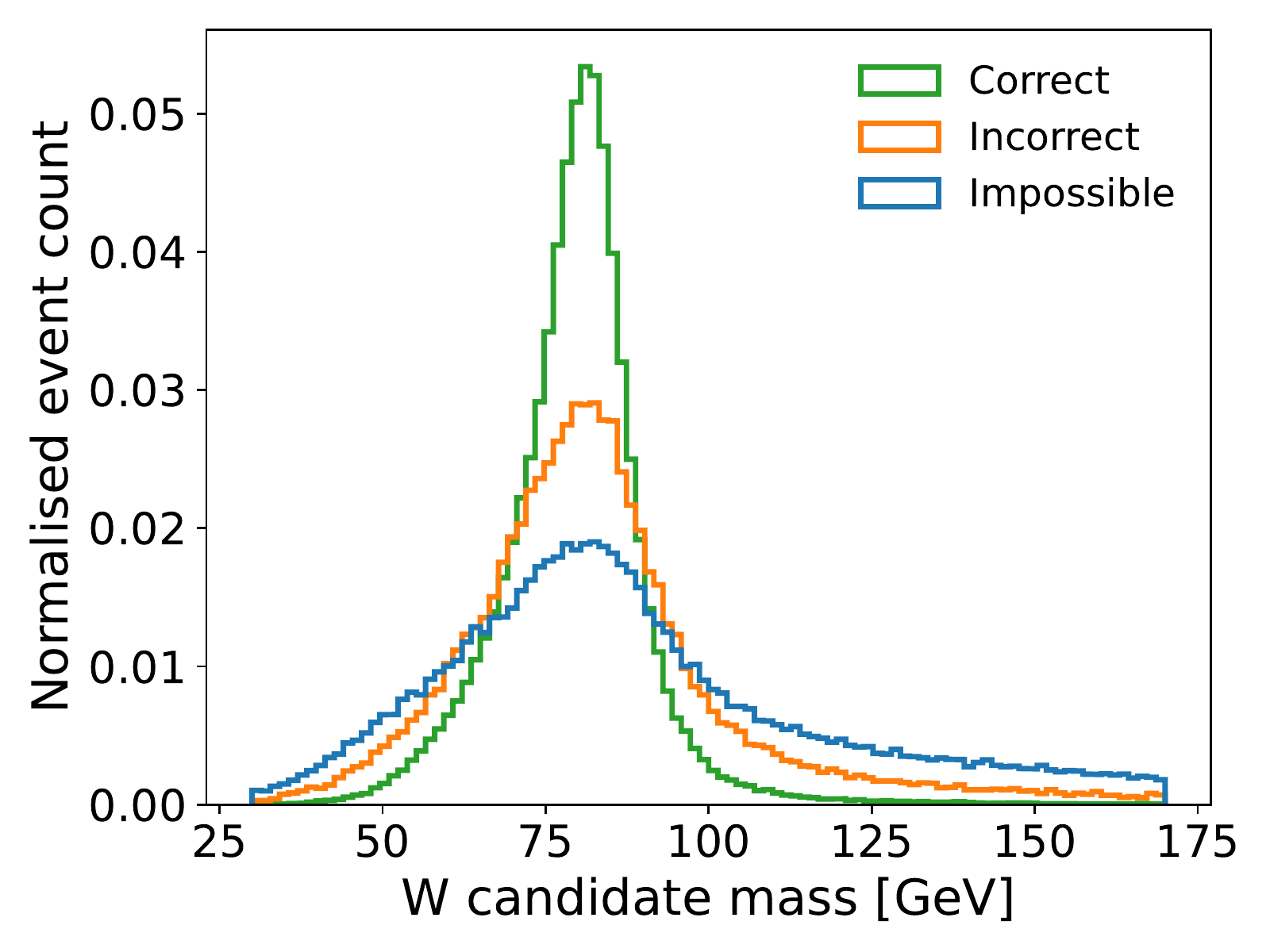}
        \caption{ }
        \label{fig:app_reco_mass_w:topo}
    \end{subfigure}
    \begin{subfigure}[b]{0.32\textwidth}
        \includegraphics[width=\textwidth]{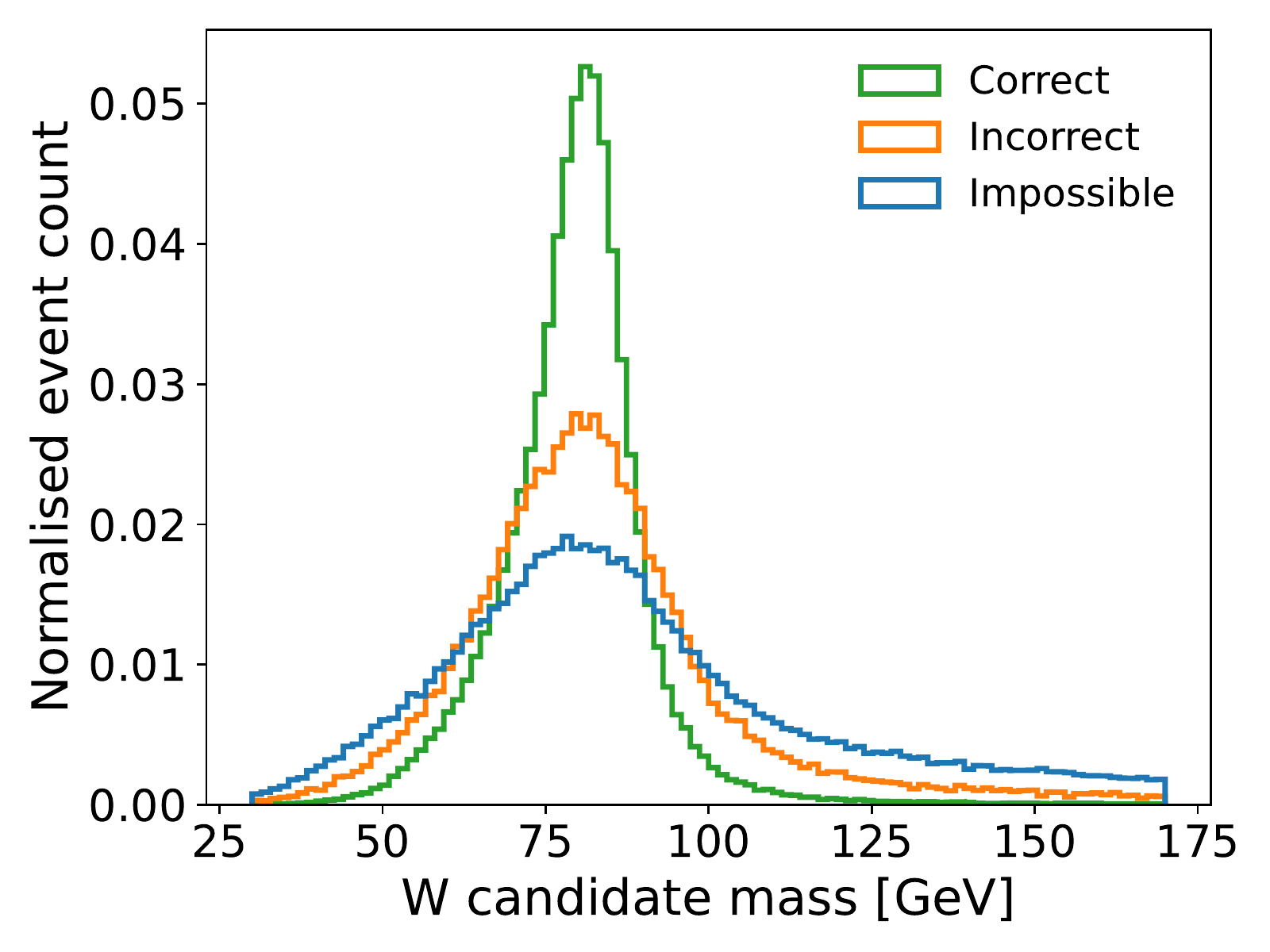}
        \caption{ }
        \label{fig:app_reco_mass_w:spa}
    \end{subfigure}
    \begin{subfigure}[b]{0.32\textwidth}
        \includegraphics[width=\textwidth]{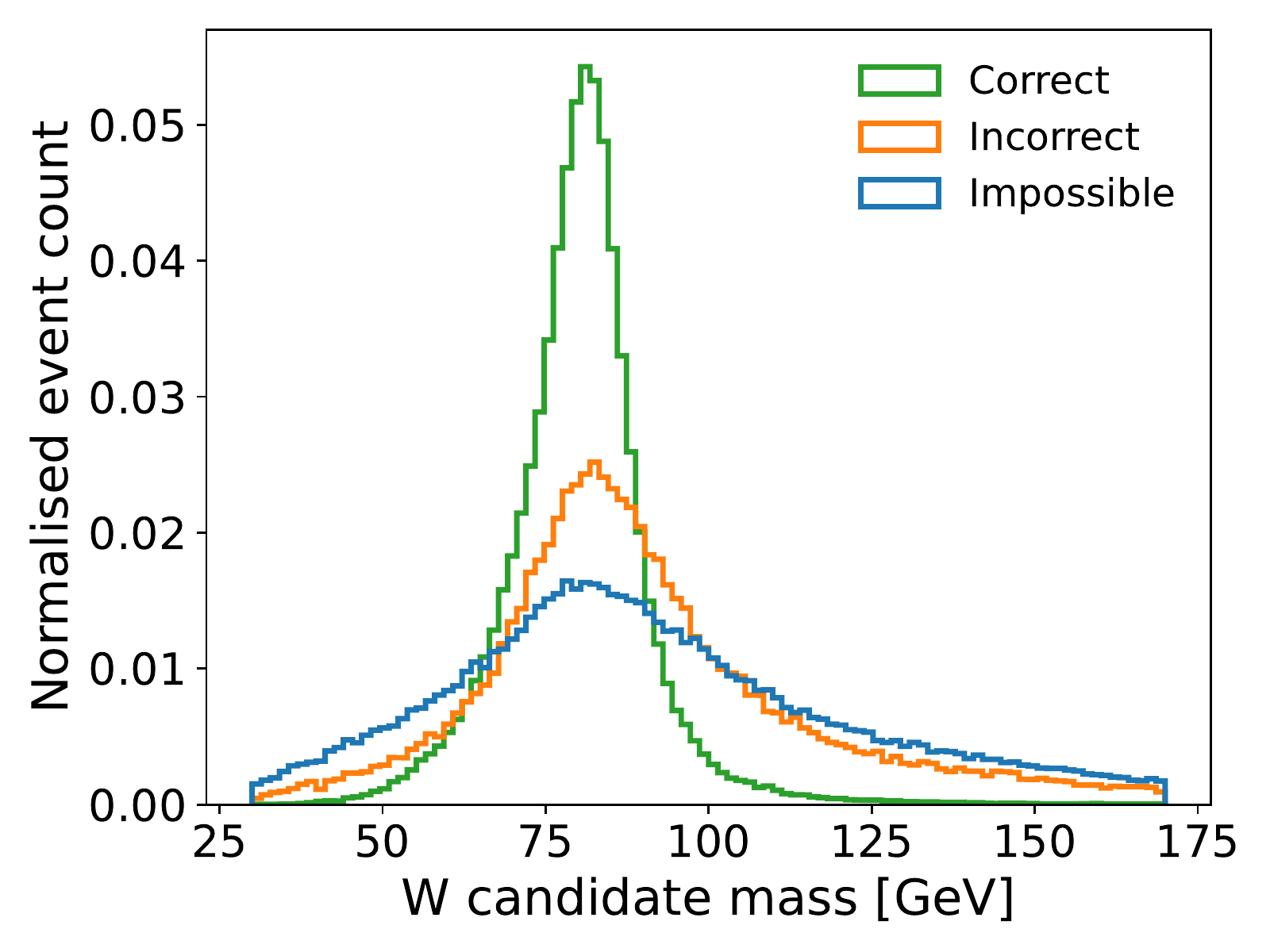}
        \caption{ }
        \label{fig:app_reco_mass_w:chi2}
    \end{subfigure} 
    \caption{Normalised distributions of the reconstructed $m_W$ using \subref{fig:app_reco_mass_w:topo} Topographs, \subref{fig:app_reco_mass_w:spa}\spanet, and \subref{fig:app_reco_mass_w:chi2} $\chi^2$.
    The single histograms split the candidates into `correct', `incorrect', and `impossible' candidates.}
    \label{fig:app_reco_mass_w}
\end{figure*}

\begin{figure*}[hbt]
    \centering
    \begin{subfigure}[b]{0.32\textwidth}
        \includegraphics[width=\textwidth]{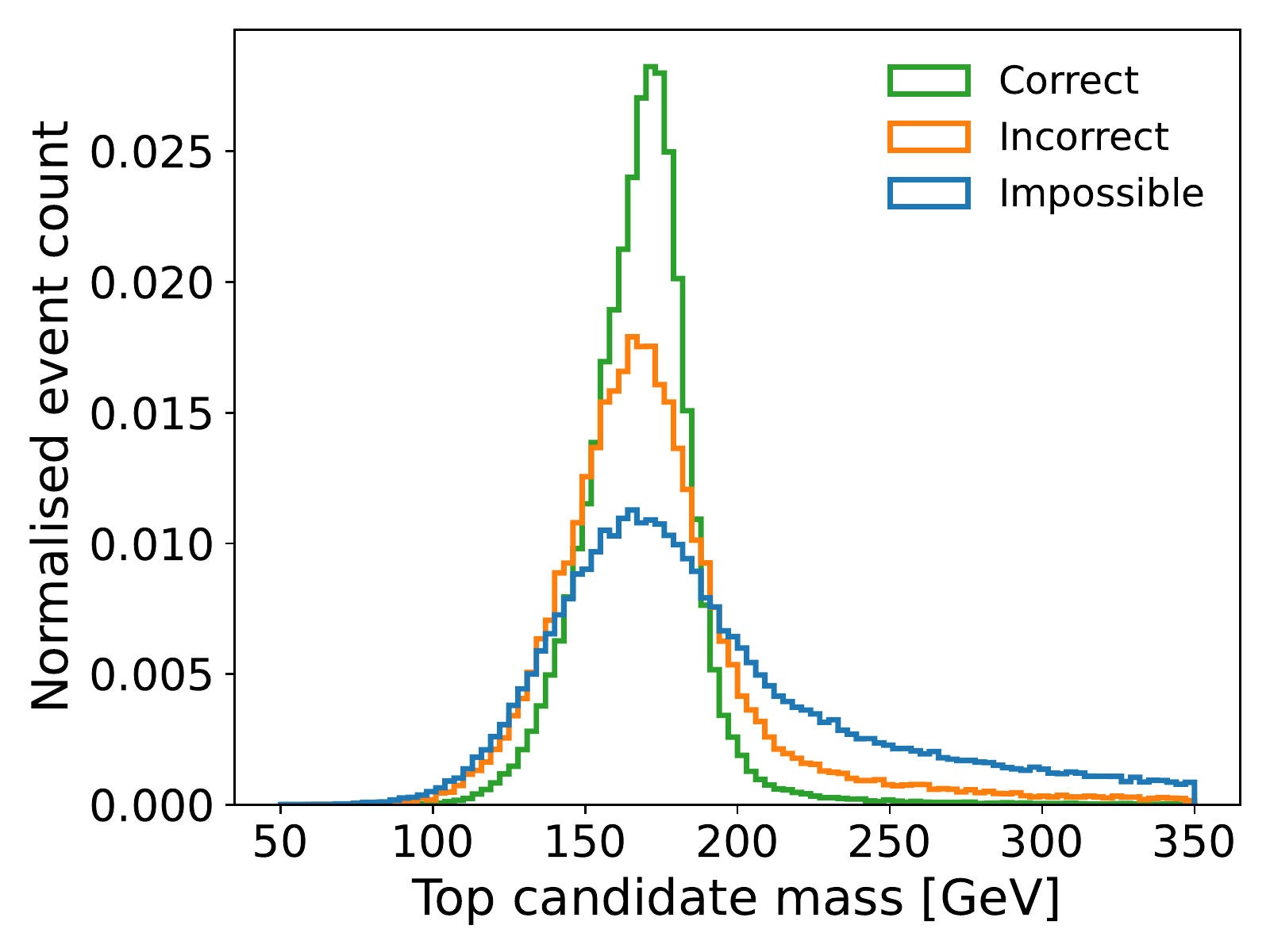}
        \caption{ }
        \label{fig:app_reco_mass_top:topo}
    \end{subfigure}
    \begin{subfigure}[b]{0.32\textwidth}
        \includegraphics[width=\textwidth]{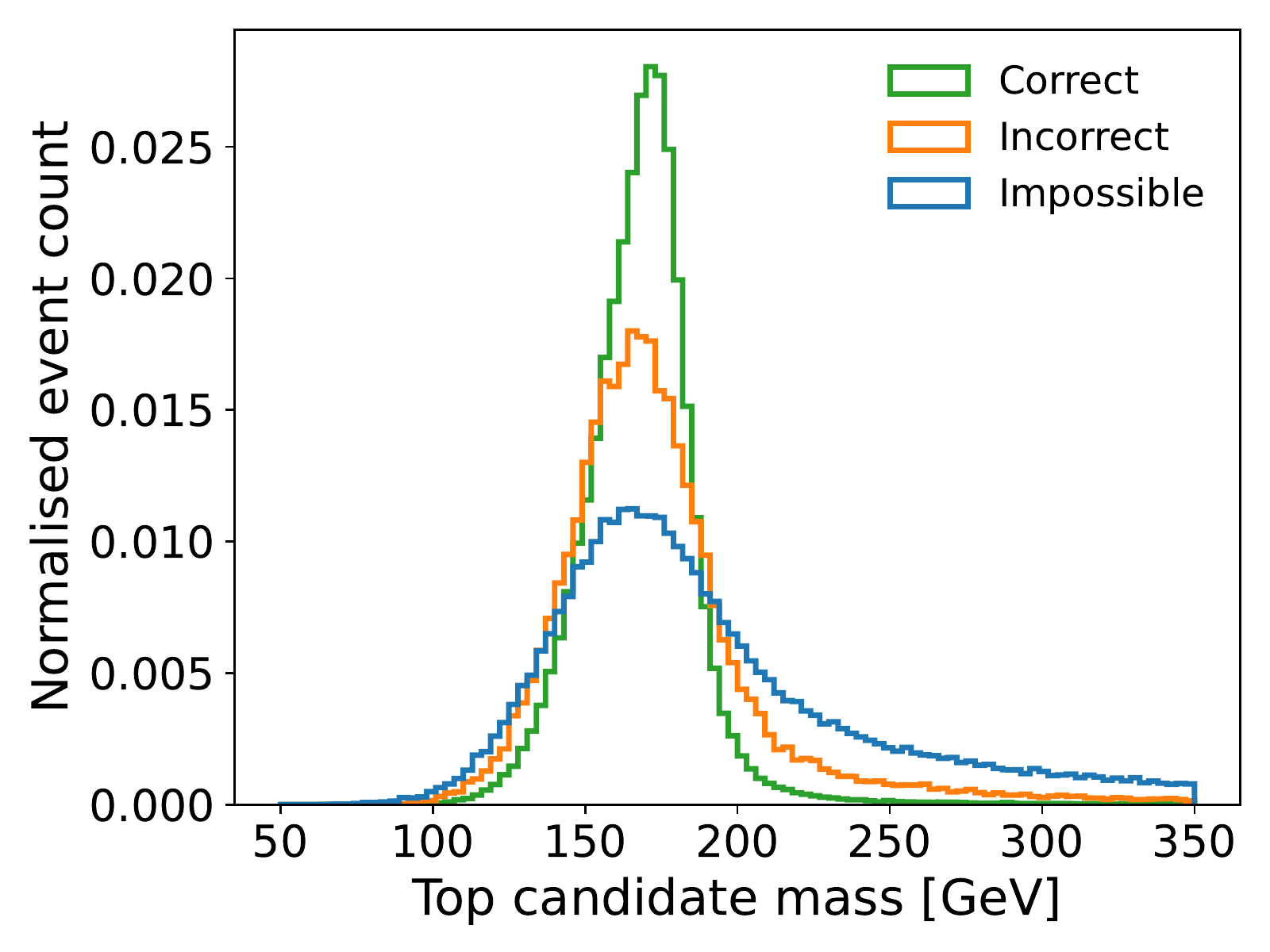}
        \caption{ }
        \label{fig:app_reco_mass_top:spa}
    \end{subfigure}
    \begin{subfigure}[b]{0.32\textwidth}
        \includegraphics[width=\textwidth]{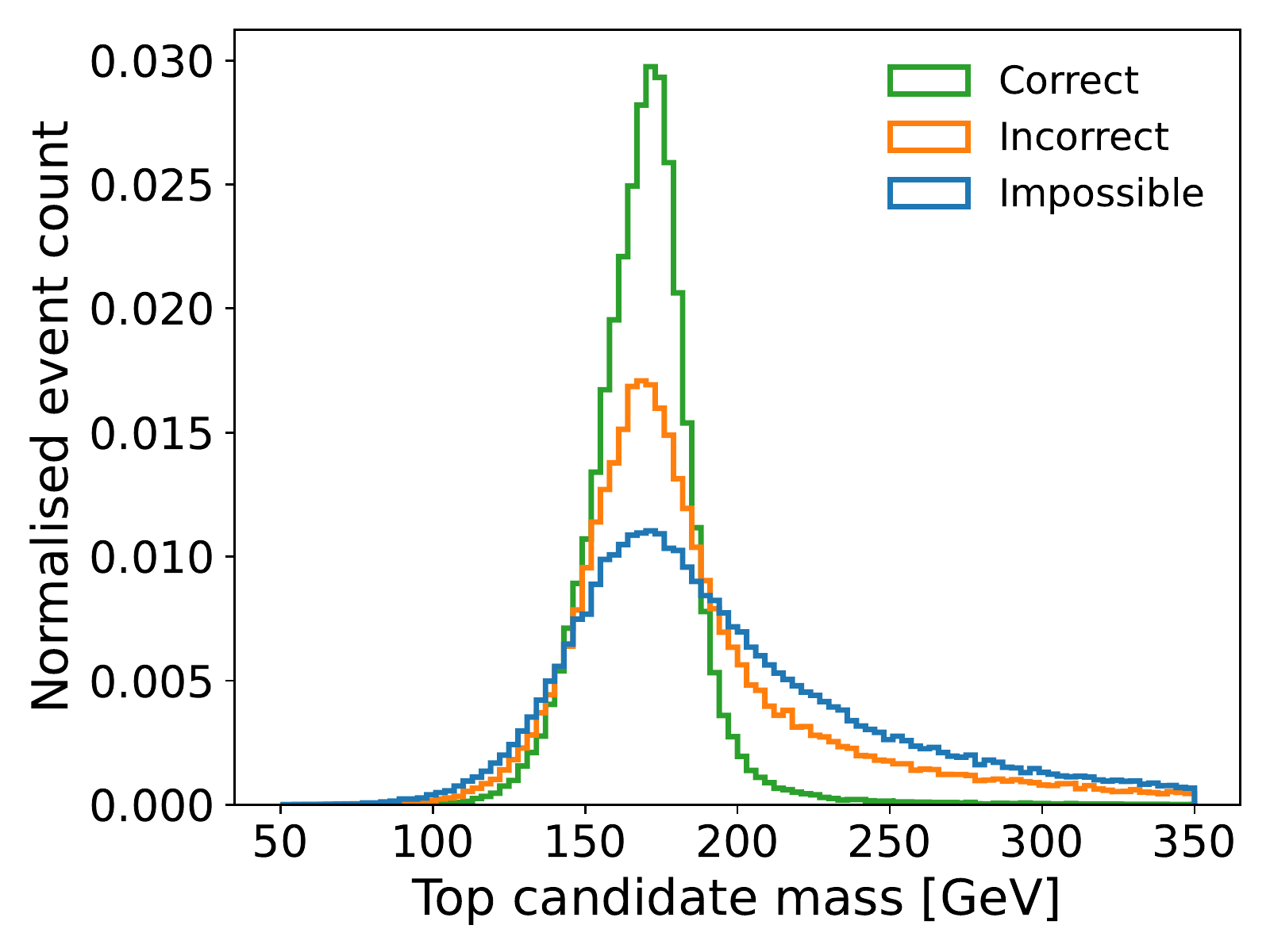}
        \caption{ }
        \label{fig:app_reco_mass_top:chi2}
    \end{subfigure} 
    \caption{Normalised distributions of the reconstructed $m_{top}$ using \subref{fig:app_reco_mass_top:topo} Topographs, \subref{fig:app_reco_mass_top:spa}\spanet, and \subref{fig:app_reco_mass_top:chi2} $\chi^2$.
    The single histograms split the candidates into `correct', `incorrect', and `impossible' candidates.}
    \label{fig:app_reco_mass_top}
\end{figure*}

\Cref{fig:app_reco_mass_w,fig:app_reco_mass_top} show the same distributions for the reconstructed $W$ and top quark mass, respectively, but every distribution is normalised to unity.

\begin{figure*}[hbt]
    \centering
    \begin{subfigure}[b]{0.45\textwidth}
        \includegraphics[width=\textwidth]{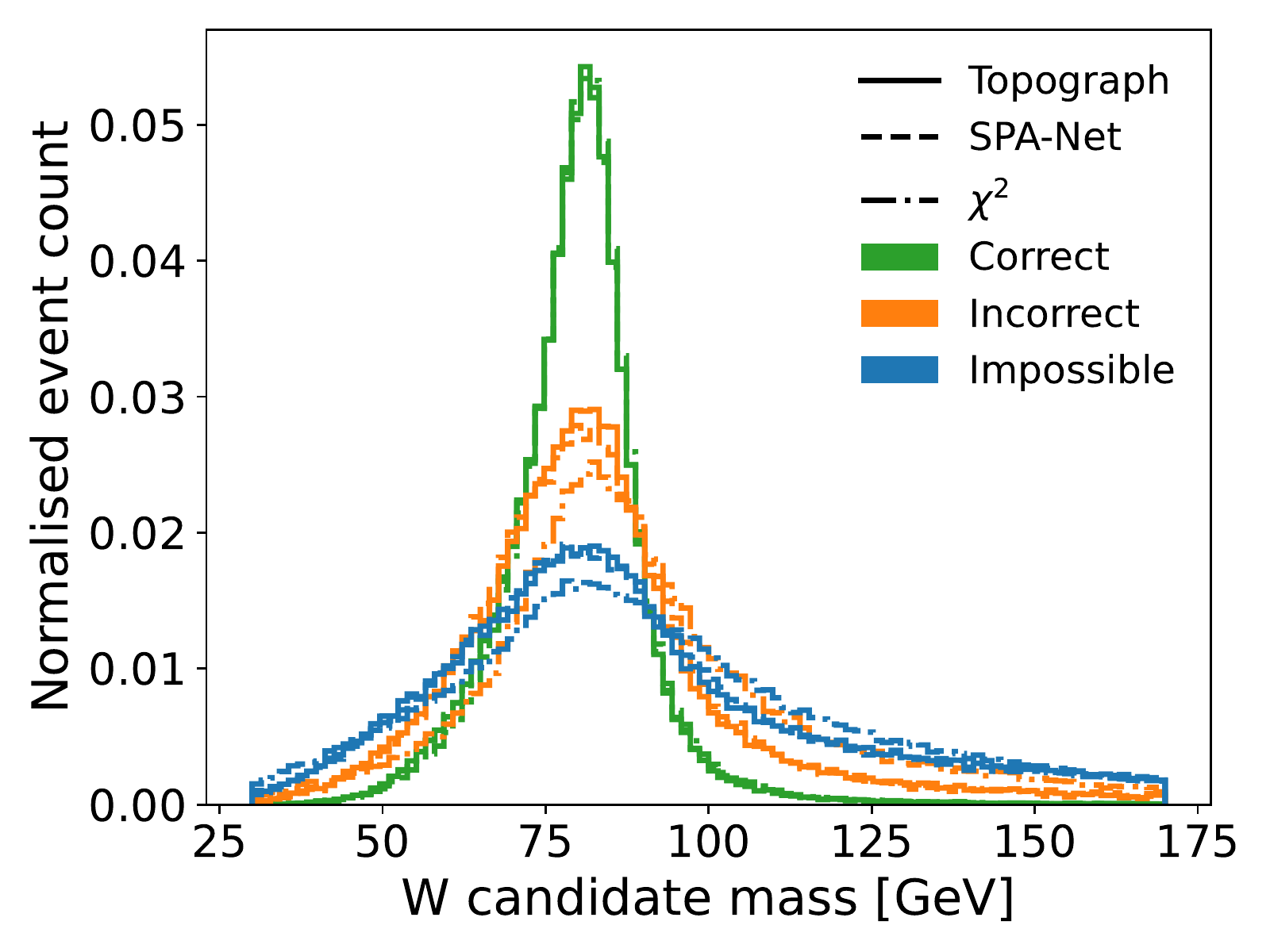}
        \caption{ }
        \label{fig:app_reco_mass_comparison_normalised:w}
    \end{subfigure}
    \begin{subfigure}[b]{0.45\textwidth}
        \includegraphics[width=\textwidth]{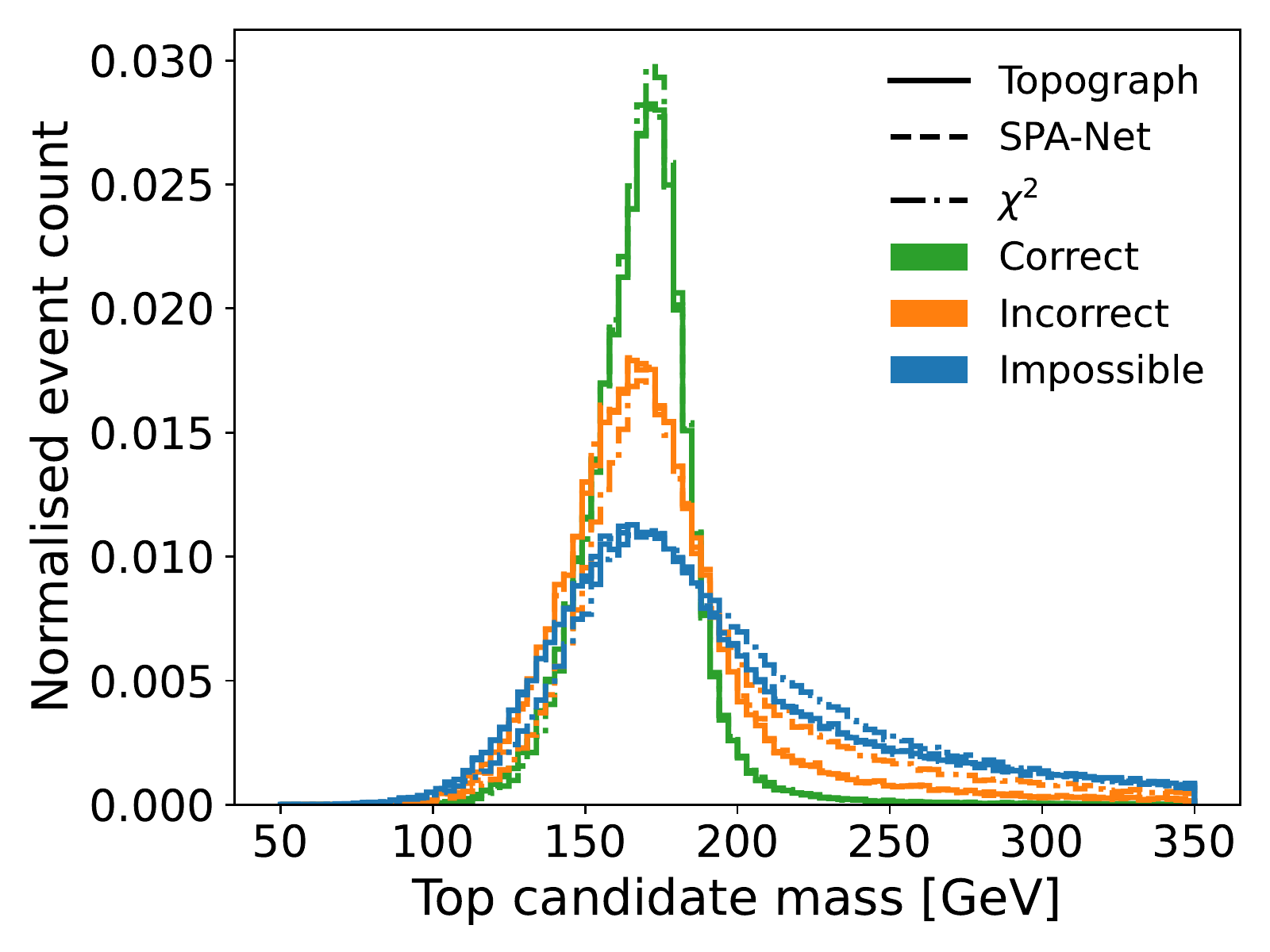}
        \caption{ }
        \label{fig:app_reco_mass_comparison_normalised:top}
    \end{subfigure}    
    \caption{Distributions of the reconstructed \subref{fig:app_reco_mass_comparison_normalised:w} $m_W$ and \subref{fig:app_reco_mass_comparison_normalised:top} $m_{top}$. 
    Each histogram is normalised to an area of one.
    The solid lines show the distributions for the Topograph, the dashed lines show the distributions for the \spanet, and the dashed-dotted lines show the distributions for the $\chi^2$ method.
    The different colours show the different types of events based on the assignment of the model: correct, incorrect, impossible.}
    \label{fig:app_reco_mass_comparison_normalised}
\end{figure*}

\Cref{fig:app_reco_mass_comparison_normalised} shows the same distributions but with the different models combined in single plots.

\end{document}